\documentclass{article}
\usepackage[lmargin=3cm,rmargin=3cm,tmargin=3.5cm,bmargin=3.5cm]{geometry}
\usepackage{amsmath, calc,amssymb}
\usepackage[main=english, polish]{babel}
\usepackage{polski}
\usepackage{graphicx} 
\usepackage{hyperref}
\usepackage{pdfcomment}
\usepackage{overpic}
\usepackage{natbib}
\usepackage{subfiles}
\usepackage{subcaption}
\usepackage{tabularx}
\usepackage{placeins}
\usepackage{xcolor}
\usepackage{authblk}
\usepackage{environ}
\usepackage{pdfcomment}
\usepackage{lineno}
\usepackage{cellspace}
\newcommand{\p}{\partial}
\newcommand{\bo}{\boldsymbol}
\newcommand{\du}{\delta u}
\renewcommand{\d}{\delta}
\newcommand{\ol}{\overline}
\newcommand{\db}{\delta b}
\newcommand{\D}{\Delta}
\newcommand{\xip}{x^{+}}
\newcommand{\xim}{x^{-}}

\newcommand{\avg}[1]{\ensuremath{\left\langle #1 \right\rangle}}
\newcommand{\mn}[1]{\ensuremath{\left\langle #1 \right\rangle}}

\title{\bf Energy cascades in  atmospheric moist convection}

\author[1]{Paweł~Jędrejko$^*$}
\author[1]{Marta~Wacławczyk}
\author[2]{Benjamin~Luce}
\author[1]{Szymon~P.~Malinowski}
\author[3]{Jakub Nowak}
\author[2]{J.~Christos~Vassilicos}

\date{
\vspace{0.3cm}
\noindent {\normalsize\textbf{Correspondence to:} Paweł Jędrejko (\texttt{p.jedrejko@uw.edu.pl})}\\\vspace{0.5cm} 
\today}

\affil[1]{Institute of Geophysics, Faculty of Physics, University of Warsaw, Warsaw, Poland}
\affil[2]{Laboratoire de Mécanique des Fluides de Lille, Univ. Lille, CNRS, ONERA, Arts et Metiers Institute of Technology, Lille, France} %
\affil[3]{Max Planck Institute for Meteorology, Bundesstraße 53, 20146 Hamburg, Germany}

\definecolor{lightskyblue}{RGB}{70, 130, 200}

\begin{document}
\maketitle

\begin{abstract}
We analyze scale-by-scale budgets of velocity and buoyancy second-order structure 
functions using Large Eddy Simulation of the BOMEX trade-wind cumulus layer. The simulation 
utilizes a $10\,\mathrm{m}$ grid resolution under the anelastic approximation, incorporating 
moist thermodynamics to capture phase changes.
The analysis of derived turbulence kinetic energy (TKE) and turbulence potential energy (TPE) 
budgets reveals four regions of distinct 
turbulence dynamics. 
For three of these regions, approximate balances holding over wide range of length scales
are identified, highlighting the importance of interspace and pressure transport. 
Above the lifting 
condensation level ($\approx 800\,\mathrm{m}$), phase changes introduce forcing across all resolved 
scales, leading to an inverse TPE cascade spanning all scales. The TKE, inter-scale transfer is 
inverse above the vertical velocity integral length scale but below the horizontal velocity scale. 
In the upper cloud layer ($>1400\,\mathrm{m}$), forward inter-scale transfers are recovered despite 
ongoing phase changes. A phenomenological interpretation of these features, referring to the 
different formation stages of convective updrafts, is also proposed. Finally, 1D horizontal subsamples from 
the simulation show good agreement with recent airborne measurements, validating the numerical 
framework and confirming the departure of moist convective turbulence from standard energy cascade models.
\vspace{0.5cm}

\end{abstract}

\section{Introduction}

Moist convection in the atmosphere refers to vertical motions of air
masses that contain water vapour and liquid water.  As warm, humid air
rises, it cools, causing the water vapour to condense and form clouds.
This phase change releases latent heat, which further amplifies the
buoyant uplift.  The intensified updraft can grow vertically,
developing into deep convective clouds and, in some cases, towering
cumulonimbus systems.

Individual convective cells do not remain isolated. They can
interact and merge into larger, organized structures --- a process
known as self-aggregation.  Through this mechanism, moist convection
becomes a key driver of thunderstorms and plays an important role in
shaping large-scale atmospheric circulation patterns
\citep{Stevens2005,Yoshiaki2013}.

Moist convection is inherently a multi-scale phenomenon.  Large
coherent thermals emerge from small-scale distributions of temperature
disturbances. Those thermals move upward transporting turbulence kinetic energy (TKE) and undergoing
self-aggregation which merges them into larger-scale dynamical structures in the atmosphere
\citep{Bony2025}. 
 
These observations may be related to the phenomenon of bidirectional
cascade known from numerous turbulence studies. Part of the energy
injected into the system through generation mechanisms subsequently
cascades towards progressively smaller eddies, ultimately dissipating
into heat via viscosity, but another part is transferred towards
larger scales allowing them to persist and grow. The dual cascades are
mostly observed in quasi-2D flow scenarios,
\citep{Smith1996,Xia2011,Byrne2011,Boffetta2023}, including the
numerical model of dry convection at mesoscales \citep{Alexakis2024}
and rotating convection \cite{Lemm2025}. In this latter work the
inverse cascade of energy, which takes place predominantly near the
bottom and top plates, follows from merging of like signed vortical
plumes. Inverse cascades were also reported in the case of 3D
turbulent convection in laboratory and numerical experiments
\citep{Togni2019,Vieweg2022,Chen2024,Lemm2025} as well as 3D helical
turbulence \cite{Plunian2020}.

In the context of atmospheric observations, \cite{Zilitinkevich2021}
argued that inverse cascades where buoyancy-generated chaotic plumes
merge to form larger plumes were an “overlooked” mechanism.  Those
authors argued that within the convective atmospheric surface layer
two scenarios are possible: a downscale transfer rate of mechanically
generated turbulence kinetic energy and an upscale transfer due to
conversion of convective turbulence kinetic energy into kinetic energy
of large-scale self-organized convective flow patterns. In this latter
case the transfer rate is equal to the transport at large scales. 
Experimental evidence of the thermal merging process in the
atmosphere was provided recently by \cite{Bony2025} who studied its
implications for clouds and circulations on larger scales.
\cite{Galina2024} focused on the role of helicity generation by
convective towers and proposed that it might be part of a possible
mechanism leading to inverse energy cascades. \cite{Yano2025} studied
the energy cycle of a convectively-organized system in an idealized
numerical model in wavelet space. They observed a tendency for kinetic
energy to be transferred horizontally to larger scales via inverse
cascade.
 
The occurrence of an inverse cascade in moist convective turbulence
could ultimately be confirmed in the statistical sense through the
analysis of a transfer rate in the space of scales. In statistically
stationary homogeneous isotropic turbulence, the third order structure
function with a negative linear scaling with separation distance
demonstrates constant (because of the linear scaling) downscale
(because of the negative sign) energy flux across scales. However, a
positive third order structure function would imply an inverse cascade
in such turbulence. For flows that are inherently non-homogeneous, the
discussion regarding the direction of energy transfer becomes more
complex. In this work we explore this problem, by studying the energy
budgets described by the K\'arm\'an–Howarth–Monin–Hill (KHMH)
equations \citep{karman1938,monin1959,hill2002}. These equations
describe how energy is transported both in the physical space and in
the space of scales. They generalize the classical K\'arm\'an–Howarth
equation to more complex situations such as inhomogeneous,
anisotropic, and/or unsteady turbulence. 
This methodology allows us to quantify how turbulence transport in 
physical space modulates the energy cascade. 
In the context of planetary-scale atmospheric flows, the KHMH equation was investigated 
by~\cite{Faranda2018}, who studied subfilter-scale energy transfers in reanalysis data, and 
by~\cite{McKinnon2026}, who showed that tropical atmospheres in a km-scale model exhibit inverse 
cascades. At smaller scales, comparable to the Atmospheric Boundary Layer (ABL) 
height,~\cite{Nowak2025} estimated the KHMH budget from airborne data measured in the trade 
cumulus regime across four different altitude levels.
Therein, it was found that in the near-surface and
middle-subcloud levels an equilibrium appears to be present for
instantaneous velocities (without Reynolds decomposition) which
appears to have similarities with classical Kolmogorov equilibrium for
fluctuating turbulent velocities but is not necessarily such an
equilibrium as it was not possible to reliably quantify the
inter-space transport rate and the turbulence production rate by mean
shear (see \cite{ALV2023} for an example of such a
situation).
The situation changed drastically at the cloud-base level, where the
third order structure function of instantaneous velocities increased
toward zero with scale. Unfortunately, due to sampling limitations of
in situ aircraft measurements the presence of an inverse cascade could
not be confirmed.

More detailed analysis can be performed with 3-dimensional numerical
data. Usually, Direct Numerical Simulations are used for this purpose.
The detailed KHMH budget was calculated for turbulent channel flow
\citep{ALV2023} shear flows \citep{ChenVassilicos2022}, turbulent jets
\citep{Lai2018} Rayleigh-B\'enard convection \citep{Togni2019} and
various other turbulent flows. A recent study of \cite{Sayeedetal2026} considered the
scale-by-scale budget of Large Eddy Simulation (LES) data.

In this work we use LES data of the moist convection in the classical
setting representing shallow cumulus convection from the BOMEX field
campaign \cite{bomexPaper}, calculated with the MicroHH code
\citep{Hee}. When applied to LES, the KHMH budget contains additional
terms due to spatial variability of the subgrid viscosity parameter.
We use the anelastic approximation, following the paper of
\cite{Bannon}. We derive the KHMH equations for the second-order
structure functions of velocity and virtual potential temperature and
analyse their budgets in the physical space and in the space of
scales. This approach is a generalisation of the common budget term
analyses of kinetic energy and temperature variance, in the sense that
apart from the vertical variation, it additionally shows how the
energy is transferred between different scales
as well as across space at given scales. This permits a structured and
systematic analysis of the interscale and interspace processes.
These processes can be related to
creation of coherent thermals and clouds and their self-aggregation. 

This paper is structured as follows. Section~\ref{secGoverningEquations} presents 
the governing equations of moist convection under the anelastic approximation. 
In Section~\ref{secSpaceScale}, we derive the scale-by-scale budgets for the 
second-order structure functions of velocity and temperature tailored to large-eddy 
simulation (LES). Section~\ref{secInputData} briefly characterizes the numerical 
setup and data, detailing the BOMEX and EUREC$^4$A configurations (cf.\ \cite{Bony2022}); 
the latter allows for direct comparison with previous analyses by \cite{Nowak2025}. 
The results are presented and discussed in Section~\ref{secResults}, followed by our 
concluding remarks in Section~\ref{secConclusions}. Finally, detailed mathematical 
derivations of the governing equations and the scale-by-scale numerical schemes are 
provided in the Appendices.

\section{Governing equations}
\label{secGoverningEquations}
In this section, we present the governing equations for dry and 
moist atmospheric convection under the anelastic framework. 
We also outline their formulation within Large Eddy Simulation 
(LES) setup, which forms the basis for our further analysis. 
We adopt a three-dimensional Cartesian coordinate system 
$\textbf{x} = (x_1,\, x_2,\, x_3) = (x,\, y,\, z)$, 
where $x$ and $y$ denote the horizontal directions and $z$ is 
the vertical coordinate (altitude). Correspondingly, the velocity 
vector is $\textbf{u} = (u_1,\, u_2,\, u_3) = (u,\, v,\, w)$.

\subsection{Basic assumptions and governing equations for dry air}

 We use the anelastic approximation, as  proposed by \cite{Bannon}.
 We assume that air is a perfect gas with the equation of state
 $p=\rho R_d T$, where $R_d$ is the specific gas constant for dry air.  
We consider the basic, reference state 
of the atmosphere, with the pressure, density and temperature
denoted, respectively as $p_0(z)$, $\rho_0(z)$ and $T_0(z)$
 and define the
density scale height $H_\rho$ 
\begin{equation}
	H_\rho =  -\left(\frac{1}{\rho_0} \frac{{\rm d} \rho_0}{{\rm d} z} \right)^{-1}.
	\label{def:H}
\end{equation}
The minus sign in the definition above indicates that $\rho_0$ decreases 
with height, ${{\rm d} \rho_0}/{{\rm d} z}<0$, as expected in the stably stratified case.
The difference between the actual  and the background state are denoted by 
$ \rho' = \rho-\rho_0$, $ p' = p-p_0$, $T'= T- T_0$.  
The anelastic approximation assumes that 
these differences (perturbations) are small compared to the static values
\begin{align}
	\frac{\rho'}{\rho_0(z)}  \ll 1, \quad 	
	\frac{p'}{p_0(z)} \ll 1, \quad 	\frac{T'}{T_0(z)} \ll 1
	\label{eq:assumption1}
\end{align}
and that the density scale height $H_\rho$ may be comparable with a displacement  of a parcel due to buoyancy.

Further assumptions  behind the anelastic approximation and derivation of the governing equations, following  \cite{Bannon}, 
are presented in detail in Appendix \ref{Appendix}.
Below  we rewrite those governing equations, cf.\  (\ref{eq:conti3})
(\ref{eq:momentum}) and (\ref{eq:energy}),
in the following form:
\begin{align}
 \frac{\partial \rho_0 u_i}{\partial x_i}  &= 0 
 \quad \text{or}
 \quad \frac{\partial u_i}{\partial x_i} = \frac{w}{H_\rho}, 
	\label{eq:anelastic_c}\\
    	 \frac{D u_i}{D t}       
    &= - \frac{\partial \pi}{\partial x_i}  
    +  g \left( \frac{\theta'}{\theta_0} \right) \delta_{i3}
    + \epsilon_{ij3} f  \left(u_j - G_j \right)
     +  \nu  \frac{\partial^2 u_i}{\partial x_j \partial x_j},   \label{eq:anelastic_m} \\
     \frac{D \theta'}{D t}  
     +  w\frac{d \theta_0}{d z}  &= \kappa \frac{\partial^2 \theta'}{\partial x_j \partial x_j}+ \frac{ \dot{Q}}{\rho_0 c_p}\left(\frac{p_{00}}{p}\right)^{R/c_p},
    \label{eq:anelastic_e}
\end{align}
where $D/Dt = \partial /\partial t + u_j \partial /\partial x_j$
$\pi = p'/\rho_0$,  $f= 2\Omega \sin(\phi)$ is the Coriolis parameter, $\phi$ is the latitude and $\textbf{G}$ is the geostrophic velocity, $\delta_{ij}$
is the Kronecker delta,  $\epsilon_{i3k}$ is the Levi-Civita symbol,
$\nu$ and $\kappa$ are the kinematic viscosity and diffusivity, respectively.
Further,
$\theta$ is the potential temperature defined as
\begin{equation}
	\theta = T \left(\frac{p_{00}}{p}\right)^{R_d/c_p},
	\label{def:potT}
\end{equation}
where $p_{00}$ is the reference pressure, $R_d= c_p - c_v$ and 
$c_p$ and $c_v$ are, respectively, the specific heat at constant pressure and at constant volume. Finally,
$\dot{Q}$ is the diabatic heating rate per unit volume.

In Eq.\ (\ref{eq:anelastic_e})
we have decomposed $\theta$ into
the sum of the background and perturbation potential temperatures 
$\theta_0 + \theta'$ assuming that 
$\kappa \partial^2 \theta_0 /\partial z^2  \ll \kappa \partial^2 \theta' /\partial x_j $, $j=1,2,3$.

The gradient of the background, potential temperature, as follows from
Eq.\ (\ref{def:potT}) and the hydrostatic balance equation, reads
\begin{equation}
	\frac{1}{\theta_0}\frac{d \theta_0(z)}{d z} = -\left(\frac{\rho_0 g}{\gamma p_0}
	+\frac{1}{\rho_0}\frac{d \rho_0}{d z}\right) := \frac{1}{g}N^2,  
\end{equation}
where $N^2$  is the buoyancy frequency.
According to \cite{Andrews}, the variations in the potential temperature $\theta$
are related to the available potential energy.
Specifically, in the limit of small perturbations the potential energy 
of dry atmosphere can be expressed as
\begin{align}
\Pi \approx \frac{1}{2 N^2} \left(g \frac{\theta'}{\theta_0}\right)^2.
\label{ep:pi}
\end{align}

\subsection{Moisture and phase changes}\label{S_moisture}

To obtain equations such as (\ref{eq:anelastic_c}--\ref{eq:anelastic_e}) for
moist convection, \cite{Bannon} expressed the density $\rho$ in terms of
the density of dry air $\rho_d$ and the mixing ratios of water vapour and liquid water,
$q_v$ and $q_l$, respectively
\begin{equation}
    \rho = \rho_d\left(1+q_v + q_l\right)
\end{equation}
As detailed in Appendix \ref{Appendix}, the momentum equation (\ref{eq:anelastic_m}) in the moist case takes the following form
\begin{equation}
    \frac{D u_i }{D  t} = -\frac{\partial \pi}{\partial x_i}  +g\left(  \frac{\theta_v'}{\theta_0} \right) \delta_{i3} +  f\epsilon_{ij3} (u_j-G_j)+\nu \frac{\partial^2 u_i}{\partial x_j \partial x_j}, \label{eq:momentum11}
\end{equation}
where $\theta_v'$ is the perturbation of the virtual potential temperature
\begin{equation}
	\theta_v = T \left(\frac{p_{00}}{p}\right)^{R_d/c_p} \left[1-\left(1-\frac{R_v}{R_d}\right)q_v - q_l \right] =
    \theta\left[1+ \epsilon q_v - q_l \right], \label{vpt}
\end{equation}
where $\epsilon = R_v/R_d - 1 \approx 0.6$.
In the moist case, in addition to the mass continuity equation, the following transport equations are solved for
$q_v$ and $q_l$
\begin{equation}
    \frac{D q_l}{D t} = C, \quad \frac{D q_v}{D t} = -C, \label{eq:ql}
\end{equation}
where $C$ is the condensation rate. In equations above, the molecular diffusion terms were neglected.

The heating rate in the energy equation (\ref{eq:anelastic_e}) is represented as
\begin{equation}
    \dot{Q} = L \rho_0  \frac{D q_l}{D t} = 
    -  L \rho_0 \frac{D q_v}{D t},
\end{equation}
where $L$ is the latent heat of vaporization and $D q_l/D t$ is equal to the condensation rate $C$, according to Eq.\ (\ref{eq:ql}). 
For further analyses, it is convenient to define
the buoyancy 
 \begin{equation}
 	b = g\frac{\theta_v'}{\theta_{0}}
 	\label{def:buoyancy}
 \end{equation}
 and  derive the transport equation for $b$. For this, 
 we first differentiate  formula (\ref{vpt})
 $$
 \frac{D \theta_v}{Dt} = \frac{D \theta}{Dt} \left[ 1+ \epsilon q_v - q_l\right]
 - \theta \frac{R_v}{R_d}  C,
 $$
 where we employed Eq.\ (\ref{eq:ql}).
 Substituting the energy equation (\ref{eq:anelastic_e}),
 after dividing by $\theta_0$ and multiplying by $g$ we obtain
\begin{align}
	\frac{D b}{D t} 
	 + w N^2  = \kappa 	\frac{\partial^2 b}{\partial x_j \partial x_j} + K C,
	\label{eq:energyb}
\end{align}
where 
$$
K = \frac{g L}{c_p T_{0}} \left[ 1+ \epsilon q_v - q_l\right] 
- \frac{R_v}{R_d}g.
$$

\subsection{Kinetic and potential energy} 

Extending approximation of the potential energy 
given by Eq.\ (\ref{ep:pi}) to moist convection,
\cite{Bannon} proposed to express it in terms of square of buoyancy
(\ref{def:buoyancy}). 
With this, the energetics of small amplitude fluctuations
are expressed via the 
 transport equation for the sum of kinetic and potential  energies derived, respectively from Eqs.\ (\ref{eq:momentum11}) and (\ref{eq:energyb})
\begin{align}
 	\frac{D}{Dt} \left( \frac{u_i u_i}{2} + \frac{b^2}{2 N^2}\right) 
  = &- 	\frac{\partial }{\partial x_i} \left(\frac{p' u_i}{\rho_0}\right) + \frac{p'w}{\rho_0 H_\rho} 
  + K C \frac{b}{N^2}  \label{eq_pot_kin_moist}\\
  &+  \frac{\partial^2 }{\partial x_j \partial x_j} \left(
  \nu \frac{u_i u_i}{2} + 
   {\kappa}\frac{b^2}{2 N^2}\right)
   - 
   \frac{\kappa}{N^2} \frac{\partial b}{\partial x_j} 
   \frac{\partial b}{\partial x_j}
   - \nu  \frac{\partial u_i}{\partial x_j}\frac{\partial u_i}{\partial x_j}, \nonumber
   \label{totalEnergyEq}
 \end{align}
provided that $N^2=\text{const}$. Terms in the second 
line of Eq.\ (\ref{eq_pot_kin_moist}) represent 
viscous diffusion and dissipation (destruction).
It can be concluded that
even  for $\nu=0$ and $\kappa=0$, 
the total energy (kinetic and potential) is not conserved
due to the pressure and source terms on the RHS
of this equation (first line).

\subsection{LES model}\label{secLES}

We consider the mass (\ref{eq:anelastic_c}), momentum (\ref{eq:momentum11})
and energy (\ref{eq:energyb}) conservation equations  subject to 
a spatial filtering procedure. We assume that the filter is homogeneous and commutes with the space derivative. The filtered quantity denoted by $\tilde{\phi}$ is calculated as the convolution
\begin{equation}
	\tilde{\phi} = \frac{1}{\Delta} \int \phi(\textbf{x}',t) G(\textbf{x}-\textbf{x}')
	\textrm{d} \textbf{x}',
\end{equation}
where $G$ is the convolution kernel.

The filtered equations read
\begin{align}
&\frac{\partial \widetilde{u_i}}{\partial x_i} 
= \frac{\tilde{w}}{H_\rho}, 
\label{conti} \\
&\frac{\tilde{D} \widetilde{u_i}}{\tilde{D} t}  
+  \frac{1}{\rho_0}  \frac{\partial \rho_0 {\tau}^a_{ij}}{\partial x_j} = -\frac{\partial \tilde{\pi}}{\partial x_i}  +  \nu \frac{\partial^2 \widetilde{u_i}}{\partial x_j \partial x_j} 
+  \delta_{i3}\tilde{b}
+ f \epsilon_{ij3} \left(\widetilde{u_j} - G_j \right),   \label{eq:Bussinesq1} \\
	&\frac{\tilde{D} \tilde{b}}{\tilde{D} t}  + \frac{1}{\rho_0}  \frac{\partial \rho_0  q_i}{\partial x_i }    = -  \tilde{w} N^2  
+ \kappa \frac{\partial^2 \widetilde{b}}{\partial x_j \partial x_j}  
+\tilde{K}\tilde{C} + f_b, 
\label{energyb}
\end{align}
where $\tilde{D}/\tilde{D} t = \partial/\partial t + \tilde{u}_j \partial/\partial x_j$  $\widetilde{u_i}$, $\tilde{b}$ and $\tilde{\pi}$ are the filtered velocity,
buoyancy and the perturbation pressure fields.
The additional source term $f_b$ in the buoyancy equation is added to account for
the unresolved large scale tendencies, including advection by subsidence velocity
(more details in~\cite{Hee}).
Additionally, the filtered source term $\widetilde{KC}$ from Eq.\ (\ref{eq:energyb}) has been modelled as
$\tilde{K} \tilde{C}$. 
The term
${\tau}^a_{ij}$ is the component of the traceless part of the 
sugbrid scale (SGS) tensor. It is modelled by the down-gradient 
eddy-viscosity assumption
$$
{\tau}^a_{ij}={\tau}_{ij} - \frac{1}{3}{\tau}_{ii} \delta_{ij}  = - 2\nu_s(\boldsymbol{x},t){S}_{ij}.
$$ 
Above,  $\nu_s$ is the SGS viscosity and
$$
{S}_{ij} = \frac{1}{2} \left( \frac{\partial \tilde{u_i}}{\partial x_j} + \frac{\partial \tilde{u_j}}{\partial x_i}
\right).
$$
The subgrid viscosity  is a function of space and time
and we will assume that $\nu_s(\boldsymbol{x},t) \gg \nu$.
 In the Smagorinsky--Lilly  model, $\nu_s$ is defined as \citep{Lilly67}
$$
\nu_s({\bf x},t) =  \lambda^2 {\cal S}\left(1-\frac{Ri}{Pr_t}\right)^{1/2}
$$
where
$$
{\cal S} = \left( 2 {S}_{ij} {S}_{ij} \right)^{1/2},
$$
and
$\lambda$ is the characteristic length scale defined as
$$
\frac{1}{\lambda^2} = \frac{1}{[\kappa(z+z_0)]^2} + \frac{1}{(c_s \Delta)^2},
$$
further, $Pr_t = 1/3$ is assumed constant and 
$$
Ri = \frac{N^2}{{\cal S}^2}.
$$
The $i$-th component of the SGS flux $q_i$ reads
$$
{q}_i = -\frac{\nu_s}{Pr_t} \frac{\partial \tilde{b}}{\partial x_i }.
$$
In Large Eddy Simulations the local rates of energy transfer to subgrid scales read \citep{Pope2000}
\begin{equation}
{\cal T}_u({\bf x},t) = -{\tau}_{ij}  {S}_{ij}. 
\qquad
{\cal T}_b({\bf x},t) = -q_i \frac{\partial \tilde{b}}{\partial x_i }.
\label{eq:T}
\end{equation}
They  approximate the energy dissipation rate and the destruction rate
of buoyancy variance 
$$
\langle {\cal T}_u({\bf x},t) \rangle \approx \varepsilon^u({\bf x},t), \qquad
\langle {\cal T}_b({\bf x},t) \rangle \approx \varepsilon^b({\bf x},t).
$$

\section{Scale-by-scale analysis methodology}
\label{secSpaceScale}

Numerous studies have analyzed the budgets of TKE 
and scalar variance within the atmospheric boundary layer 
\citep{Stull1988,Kaimaletal1976}. These budgets elucidate how energy is produced, 
transported, and dissipated across different altitudes, providing the essential 
closures  ("parameterizations") required for numerical weather prediction and 
climate modeling.

However, atmospheric turbulence is an inherently multiscale phenomenon, structured 
as a continuous hierarchy of motions. At the largest scale, geostrophic forcing 
and surface friction establish a prominent, sheared mean horizontal flow. This macro-scale 
shear, combined with surface heating, feeds energy into organized, boundary-layer-spanning 
convective structures, such as horizontal convective rolls and rising buoyant plumes. 
These large structures eventually break down into a chaotic sea of smaller, multi-dimensional 
turbulent fluctuations. Consequently, energy is not only transported through physical 
space by these distinct structural layers (inhomogeneity), but it is also dynamically 
transferred between eddies of varying sizes. 
A critical question
is to know at which scales the interscale energy transfer in moist convection is forward or 
backward, and how either occurrence may correlate with interspace transfers at given scales. 

\subsection{Reynolds decomposition}

To formulate the scale-by-scale budget equations, we first apply a standard 
Reynolds decomposition to separate the large-scale mean background flow from 
the resolved fluctuations. For any arbitrary resolved field $\phi$, the 
horizontal- and time-averaged mean component $\langle\phi\rangle$ is denoted
with a shorthand notation:
\begin{equation*}
    \mn{\phi} = \langle \phi \rangle_{xyt}
\end{equation*}
where the corresponding fluctuating component is denoted by a prime:
\begin{equation*}
    \phi' = \phi - \mn{\phi}. 
\end{equation*}
Note that the prime notation $'$ is used in two distinct contexts in this work. In 
Section \ref{secGoverningEquations}, it denotes perturbations from the hydrostatic reference 
state. 
By defining the mean state strictly through domain-wide horizontal and temporal averaging, 
the resulting fluctuating component $\phi'$ captures the entire spectrum of motions 
superimposed on the background wind and thermodynamic profiles. This includes both the 
highly chaotic, small-scale motions and the deeper, organized convective turbulent 
structures such as buoyant plumes. 

In the context of atmospheric modeling, combining these features into a single fluctuating 
pool is physically justified. Both the convective-scale updrafts and the small-scale 
fluctuations occupy the subgrid spectrum of coarse weather prediction and climate models, 
where they are collectively parameterized rather than explicitly resolved. Our scale-by-scale 
approach treats them as a continuous spectrum of fluctuating energy whose scale-dependent 
transfers can be tracked continuously through the exact budget equations.

Applying this Reynolds decomposition to the implicitly filtered LES momentum equation 
(\ref{eq:Bussinesq1}) yields the following transport equation for the velocity fluctuations (the tilde $\tilde{\cdot}$ notation is omitted for brevity):
\begin{equation}
    \frac{\p u_i'}{\p t} = 
    -\frac{\p u_i'}{\p x_j} u_j' 
    -\frac{\p u_i'}{\p x_j} \langle u_j \rangle 
    -\frac{\p \langle u_i \rangle}{\p x_j} u_j' 
    -\frac{\p\pi'}{\p x_i}
    +\frac{1}{\rho_0}\frac{\p}{\p x_j} \left(2\nu_s\rho_0 S_{ij}'\right) 
    + \delta_{i3}b' 
    + f\epsilon_{ij3}u_j' + M_u
    \label{uFluctEq}
\end{equation}
and buoyancy fluctuations derived from Eq.\ (\ref{energyb}):
\begin{equation}
    \frac{\p b'}{\p t} = 
    -\frac{\p b'}{\p x_j} u_j' 
    -\frac{\p b'}{\p x_j} \langle u_j \rangle 
    -\frac{\p \langle b \rangle}{\p x_j} u_j' 
    +\frac{1}{\rho_0}\frac{\p}{\p x_j} \left(\kappa_s\rho_0 \frac{\p b'}{\p x_j}\right) 
    - w'N^2
    + (KC)' + f_b' + M_b
    \label{bFluctEq}
\end{equation}
where $M_u$ and $M_b$ denote purely mean terms, which are not detailed here for brevity as they vanish 
upon subsequent multiplication by a fluctuating quantity and ensemble averaging. 
To summarize, equations (\ref{uFluctEq}) and (\ref{bFluctEq}) govern the fields that have 
been first implicitly filtered by the LES grid and subsequently separated into mean and 
fluctuating components via the spatio-temporal Reynolds operator.

\subsection{Scale-by-scale operators}

To track the transfer of fluctuating energy across both physical positions 
$\bo{X} = (X_1, X_2, X_3) = (X, Y, Z)$ and spatial separation scales 
$\bo{r} = (r_1, r_2, r_3) = (r_x, r_y, r_z)$, 
we introduce a two-point 
coordinate transformation. 
Let a pair of distinct points in physical space ($x$-space) be denoted by 
$\bo{x}^-$ and $\bo{x}^+$. Similarly, a field $\phi$ evaluated at these 
points is denoted as $\phi^-$ and $\phi^+$, respectively. These physical 
coordinates are mapped into a combined position-separation space 
($\bo{r}\text{--}\bo{X}$ space) via:
\begin{equation*}
    X_i = \frac{{x}^+_i + {x}^-_i}{2}, \qquad\qquad r_i = \frac{{x}^+_i - {x}^-_i}{2}
\end{equation*}
which yields the inverse transformation:
\begin{equation}
    {x}^-_i = X_i - r_i, \qquad\qquad {x}^+_i = X_i + r_i.
    \label{xiTorX}
\end{equation}
In a statistically homogeneous framework, the flow statistics become independent 
of the centroid position $X_i$, and if the system exhibits statistical 
isotropy, they depend solely on the scalar modulus $r=|\bo{r}|$. 

When formulating the scale-by-scale transport equations for the second-order 
structure functions, it is mathematically advantageous to define a two-point 
increment operator ($\d$) and a two-point average operator ($\ol{\cdot}$):
\begin{equation}
    \d\phi = \phi^+ - \phi^-,
    \qquad\qquad
    \ol{\phi} = \frac{\phi^+ + \phi^-}{2}.
    \label{deltaBarOperators}
\end{equation}
These operators can be shown to satisfy 
the following exact algebraic identities:
\begin{subequations}
    \begin{equation}
        \d(\phi\psi) = \d\phi\,\ol{\psi} + \ol{\phi}\,\d\psi
        \label{deltaOfProd}
    \end{equation}
    \begin{equation}
        \ol{\phi\psi} = \tfrac{1}{4}\d\phi\,\d\psi + \ol{\phi}\,\ol{\psi}
        \label{barOfProd}
    \end{equation}
    \begin{equation}
        \d\frac{\p\phi}{\p x_i} = \frac{\partial \phi^+}{\partial {x}^+_i} - 
        \frac{\partial \phi^-}{\partial {x}^-_i}
        = \frac{\p\d\phi}{\p X_i} = 2\frac{\p \ol{\phi}}{\p r_i}
        \label{deltaOfGrad}
    \end{equation}
    \begin{equation}
        \ol{\frac{\p\phi}{\p x_i}} = \frac{1}{2} \left(
        \frac{\partial \phi^+}{\partial {x}^+_i} + 
        \frac{\partial \phi^-}{\partial {x}^-_i}
        \right) =
         \frac{\p\ol{\phi}}{\p X_i} = \frac{1}{2}\frac{\p\d\phi}{\p r_i}.
        \label{barOfGrad}
    \end{equation}
\end{subequations}
Applying these operators on the resolved anelastic continuity equation (\ref{conti}) 
yields two constraints governing the velocity increments and mid-point averages:
\begin{subequations}
    \begin{equation}
        \frac{\p\d u_i}{\p X_i} = 2\frac{\p\ol{u}_i}{\p r_i} = \d\Big(\frac{u_3}{H_\rho}\Big),
        \label{deltaOfCont}
    \end{equation}
    \begin{equation}
        \frac{\p\ol{u}_i}{\p X_i} = \frac{1}{2}\frac{\p\d{u}_i}{\p r_i} = \ol{\Big(\frac{u_3}{H_\rho}\Big)}.
        \label{barOfCont}
    \end{equation}
\end{subequations}

\subsection{Scale-by-scale budget formulation}

The overall strategy for the scale-by-scale budget analysis is to isolate 
the governing equation for the fluctuating components, derive the transport 
equation for the spatial increment $\d \phi'$, and then multiply the result 
by $2\d\phi'$. 
Symbolically, this operation can be expressed as:
\begin{equation}
    \d\big[\text{governing equation}(\phi')\big]\cdot 2\d\phi'.
    \label{spaceScaleGov}
\end{equation}
Finally, we apply an ensemble average, which is numerically approximated by 
integrating the fields over time and the horizontal homogeneous coordinates:
\begin{equation*}
    \langle\d\big[\text{governing equation}(\phi')\big]\cdot 2\d\phi'\rangle = 
    \frac{1}{T L_x L_y}\int_0^T \int_0^{L_y} \int_0^{L_x} \d\big[\text{governing equation}(\phi')\big]\cdot 2\d\phi' \, dXdYdt.
\end{equation*}

We illustrate the details using the nonlinear turbulent 
advection term of the fluctuating velocity equation (\ref{uFluctEq}) 
as an example. 
First, we apply $\delta$ operator and dot the result with $2\du_i'$, 
following directly (\ref{spaceScaleGov}). Then we use 
the algebraic identity \eqref{deltaOfProd}:
\begin{equation*}
    2\d\Big(\frac{\p u_i'}{\p x_j}u_j'\Big)\d u_i' 
    = 
    2\Big(\d\frac{\p u_i'}{\p x_j}\ol{u}_j' + \ol{\frac{\p u_i'}{\p x_j}}\d{u}_j'\Big)\d u_i'.
\end{equation*}
We further use the formulas \eqref{deltaOfGrad} and \eqref{barOfGrad}, 
apply the chain rule, and invoke 
the anelastic continuity relation \eqref{barOfCont}. Finally, the ensemble-averaged 
term becomes:
\begin{equation*}
    \left\langle\frac{\p\d u_i'^2}{\p X_j}\ol{u}_j' + \frac{1}{2}\frac{\p \d u_i'^2}{\p r_j}\d{u}_j' \right\rangle
    =
    T^u_u + S_3^u + A'^u
\end{equation*}
where the resulting budget contributions are defined as:
\begin{align}
    T^u_u &=  \left\langle \frac{\p\d u'^2\ol{u}_j' }{\p X_j} \right\rangle,
    \\
    S^u_3 &=  \frac{1}{2}\left\langle\frac{\p \d u'^2\d{u}_j' }{\p r_j} \right\rangle,
    \\
    A'^u &= -2 \left\langle  \ol{\Big(\frac{u_3'}{H_\rho}\Big)}\d u'^2 \right\rangle.
\end{align}
Physically, $T^u_u$ represents the scale-by-scale transport of TKE 
in $X$-space (i.e., among structures of the same size but at different 
locations) mediated by the turbulent fluctuations themselves. 
Conversely, $S_3^u$ represents the analogous TKE transfer in $r$-space 
(i.e., among structures at the same position but of different sizes). 
The remaining 
term, $A'^u$, represents the specific contribution arising from the 
anelastic approximation.

The other two advective terms in Eq.~\eqref{uFluctEq} are treated 
in a similar fashion. This procedure yields $\Pi_X^u$ and $\Pi_r^u$, 
which represent the analogous processes executed 
instead by the mean background flow. It also yields $P_X^u$ and $P_r^u$, 
which represent TKE production via two-point Reynolds stresses acting 
on the mean flow deformation tensor. All corresponding anelastic 
residuals are combined into a single unified term, $A^u$. The complete 
set of precise formulations for these terms is compiled in Table~\ref{termsTab}.

As a second illustrative example, we outline how the subgrid-scale 
diffusive terms are treated. Referring to Eq.~\eqref{uFluctEq}, we split 
the divergence via the product rule:
\begin{equation*}
    \frac{1}{\rho_0}\frac{\p}{\p x_j}\Big(2\rho_0\nu_s S'_{ij} \Big)
    =
    \frac{\p}{\p x_j}\Big(2\nu_s S'_{ij} \Big) 
    -
    \nu_s S'_{ij}\frac{1}{H_\rho}
\end{equation*}
to isolate the anelastic contribution into a separate term. We now focus 
on the first term on the right-hand side, following the strategy outlined 
in \ref{spaceScaleGov}. First, identity \eqref{deltaOfGrad} is applied 
to split the spatial gradient into its $\bo{X}$ and $\bo{r}$ components equally:
\begin{equation*}
 4\d\frac{\p}{\p x_j}\Big(  \nu_s S'_{ij} \Big) \du_i 
 =
 2\frac{\p}{\p X_j}\d\Big(\nu_s S'_{ij}\Big) \du_i
 +
 4\frac{\p}{\p r_j}\ol{\nu_s S'_{ij}} \du_i.
\end{equation*}
Next, we invoke the algebraic identities \eqref{deltaOfProd} and \eqref{barOfProd}:
\begin{equation*}
 =
 2\frac{\p}{\p X_j}
 \Big(
 \d\nu_s \ol{S'_{ij}}
 + 
 \ol{\nu}_s\d S'_{ij}
 \Big) 
 \du_i    
 +
 \frac{\p}{\p r_j}
 \Big(
 \d\nu_s\d S'_{ij}
 +
 4\ol{\nu}_s\ol{S'_{ij}}
 \Big) 
 \du_i,
\end{equation*}
followed by the gradient relations \eqref{deltaOfGrad} and \eqref{barOfGrad}:
\begin{equation*}
 =
 \frac{\p}{\p X_j}
 \Big(
 \d\nu_s S_{ij}^{r,\d}
 +
 2\ol{\nu}_s S_{ij}^{X, \d }
 \Big) 
 \d u_i
 +
 \frac{\p}{\p r_j}
 \Big(
 \d\nu_s S_{ij}^{X,\d}
 +
 2\ol{\nu}_s S_{ij}^{r,\d}
 \Big) 
 \d u_i,
\end{equation*}
and finally, applying the product rule and ensemble averaging
we obtain:
\begin{equation*}
     D_{Xr} + D_{XX} + D_{rX} + D_{rr} - 4\varepsilon
\end{equation*}
\begin{equation}
    D_{XX} = 2 \avg{ \frac{\p}{\p X_j} \ol{\nu}_s S_{ij}^{X,\d}\du_i
    },
    \qquad
    D_{Xr} = \avg{ \frac{\p}{\p X_j} \d\nu_s S_{ij}^{r,\d}\du_i }
\end{equation}
\begin{equation*}
    D_{rX} = \avg{\frac{\p}{\p r_j}\d\nu_s S_{ij}^{X,\d}\du_i },
    \qquad
    D_{rr} = 2 \avg{\frac{\p}{\p r_j} \ol{\nu}_s S_{ij}^{r,\d}\du_i },
\end{equation*}
\begin{equation*}
    4\varepsilon = 8  \avg{\ol{\nu_s S_{ij}' S_{ij}'} },
\end{equation*}
where
\begin{equation*}
    S_{ij}^{X, \d} = \frac{1}{2}\Big(\frac{\p\du_j'}{\p X_i}
    +\frac{\p\du_i'}{\p X_j} \Big),
    \qquad 
    S_{ij}^{r, \d} = \frac{1}{2}\Big(\frac{\p\du_j'}{\p r_i}
    +\frac{\p\du_i'}{\p r_j} \Big).
\end{equation*}

\subsection{Summary of terms}
\label{secTerms}
The derivation of the other terms is omitted for brevity.
These include the transient term $t^u$, buoyancy work term $W^u$, and the total
pressure transport $T_p^u$ (which is slightly perturbed by the anelastic contribution). 
The terms in the buoyancy equation are derived analogously.

Both -- the velocity ($^u$) and buoyancy ($^b$) KHMH equations 
(the latter derived from Eq.\ (\ref{bFluctEq})), are summarized below with 
term definitions gathered in Table (\ref{termsTab}).
Their values are computed numerically, as described in Appendix \ref{secNumericalMethod}.

\begin{subequations}
    \begin{equation}
        t^u = - S_3^u -\Pi_r^u +P_r^u - T_u^u -\Pi_X^u + P_X^u - A^u + D_{XX}^u + 
        D_{Xr}^u + D_{rX}^u + D_{rr}^u + D_A^u - 4\varepsilon^u + W^u + T_p^u
        \label{uSummary}
    \end{equation}
    \begin{equation}
        t^b = - S_3^b -\Pi_r^b +P_r^b - T_u^b -\Pi_X^b + P_X^b - A^b + D_{XX}^b + 
        D_{Xr}^b + D_{rX}^b + D_{rr}^b + D_A^b - 4\varepsilon^b + W^b + Q^b
        \label{bSummary}
    \end{equation}
    \label{KHMHeq}
\end{subequations}
Note that only 4 terms, i.e.\ the transport  in physical space $T^u$,
the transfer rate $S_3^u$, the buoyancy work $W^u$ and the dissipation
$-4\varepsilon$ were considered in \cite{Nowak2025}. The diffusion terms denoted
with symbol $D$ appear due to the subgrid scale contributions in LES model. In
particular, the mixed terms $D^u_{rX}$, $D^u_{Xr}$, $D^b_{rX}$ and $D^b_{Xr}$,
appear due to the spatial variability of the SGS viscosity $\nu_s$.

\begin{table}[hb!]
\centering
\begin{subtable}[t]{0.45\textwidth}
\centering
\renewcommand{\arraystretch}{1.8}
\begin{tabular}{||Sl|Sl||}
     \hline
     term & formula \\
     \hline \hline
     $\phantom{-}t^u$ & $\avg{ \frac{\p \du'^2}{\p t} }$ \\ \hline
     $-S_3^u$ & $-\frac{1}{2}\avg{\frac{\p \du'^2\du'_j}{\p r_j}}$ \\ \hline
     $-\Pi_r^u$ & $-\frac{1}{2}\avg{\frac{\p{\du'}^2\delta \mn{u}_j}{\p r_j}}\phantom{\bigg|}$ \\ \hline
     $\phantom{-}P_r^u$ & $-\frac{1}{2}\avg{\du_i'\du_j'\frac{\p\delta \mn{u}_i}{\p r_j}}$ \\ \hline
     $-T_u^u$ & $-\avg{\frac{\p \du'^2\ol{u}'_j}{\p X_j}}$ \\ \hline
     $-\Pi_X^u$ & $-\avg{\frac{\p{\du'}^2 \ol{\mn{u}}_j}{\p X_j}}$ \\ \hline
     $\phantom{-}P_X^u$ & $-\avg{\du_i'\ol{u}_j'\frac{\p \delta \mn{u}_i}{\p X_j}}$ \\ \hline
     $-A^u$ & $2\avg{\ol{\frac{u_3}{H_\rho}}\du'^2}$  \\ \hline
     $D_{XX}^u$ & $2\avg{\frac{\p}{\p X_j}\Big(\ol{\nu}_s S_{ij}^{X,\d}\du_i'\Big)  }$ \\ \hline
     $D_{Xr}^u$ & $ \avg{\frac{\p}{\p X_j}\Big(\d{\nu}_s S_{ij}^{r,\d}\du_i'\Big)  }$ \\ \hline
     $D_{rX}^u$ & $ \avg{\frac{\p}{\p r_j}\Big(\d{\nu}_s S_{ij}^{X,\d}\du_i'\Big)  }$ \\ \hline
     $D_{rr}^u$ & $2\avg{\frac{\p}{\p r_j}\Big(\ol{\nu}_s S_{ij}^{r,\d}\du_i'\Big)  }$ \\ \hline
     $D_{A}^u$ & $4\avg{\d\Big(\frac{\nu_s}{H_\rho}S_{i3}\Big)\du_i} $ \\ \hline
     $4\varepsilon^u$ & $8\avg{\ol{\nu_s S_{ij}'S_{ij}'}}$ \\ \hline
     $W^u$ & $2\avg{\db'\du_3'}$ \\ \hline
     $T_p^u$ & $-2\avg{\frac{\p\d \pi'}{\p X_i}\du_i'}$ \\ %
     \hline
\end{tabular}
\caption{velocity second order structure function}
\end{subtable}
\hfill
\begin{subtable}[t]{0.45\textwidth}
\centering
\renewcommand{\arraystretch}{1.8}
\begin{tabular}{||Sl|Sl||}
     \hline
     term & formula \\
     \hline \hline
     $\phantom{-}t^b$ & $\avg{  \frac{\p \db'^2}{\p t}  }$ \\ \hline
     $-S_3^b$ & $-\frac{1}{2}\avg{ \frac{\p \db'^2\du_j'}{\p r_j}  }$ \\ \hline
     $-\Pi_r^b$ & $-\frac{1}{2}\avg{\frac{\p{\db'}^2\delta \mn{u_j}}{\p r_j}} \phantom{\bigg|}$ \\ \hline
     $\phantom{-}P_r^b$ & $-\frac{1}{2}\avg{\db'\du_j'\frac{\p\delta \mn{b}}{\p r_j}}$ \\ \hline
     $-T_u^b$ & $-\avg{ \frac{\p \db'^2\ol{u}_j'}{\p X_j}  }$ \\ \hline
     $-\Pi_X^b$ & $-\avg{\frac{\p{\db'}^2 \ol{\mn{u}}_j}{\p X_j}}$ \\ \hline
     $\phantom{-}P_X^b$ & $-\avg{\db'\ol{u}_j'\frac{\p \delta \mn{b}}{\p X_j}}$ \\ \hline
     $-A^b$ & $2\avg{\ol{\frac{u_3}{H_\rho}}\db'^2 }$  \\ \hline
     $D_{XX}^b$ & $2\avg{\frac{\p}{\p X_j}\Big(\ol{\kappa}_s \frac{\p \db'^2}{\p X_j}\Big)  }$ \\ \hline
     $D_{Xr}^b$ & $ \avg{\frac{\p}{\p X_j}\Big(\d{\kappa}_s \frac{\p \db'^2}{\p r_j}\Big)  }$ \\ \hline
     $D_{rX}^b$ & $ \avg{\frac{\p}{\p r_j}\Big(\d{\kappa}_s \frac{\p \db'^2}{\p X_j}\Big)  }$ \\ \hline
     $D_{rr}^b$ & $ 2\avg{\frac{\p}{\p r_j}\Big(\ol{\kappa}_s \frac{\p \db'^2}{\p r_j}\Big)  }$ \\ \hline
     $D_{A}^b$ & $4\avg{\d\Big(\frac{\kappa_s}{H_\rho}\frac{\p \db'^2}{\p \xi_3}\Big)\du_i' }$ \\ \hline
     $4\varepsilon^b$ & $8\avg{\ol{\kappa_s \frac{\p \db'}{\p \xi_j}\frac{\p \db'}{\p \xi_j} } }$ \\ \hline
     $W^b$ & $-2\avg{\db'\d(u_3' N^2)}$ \\ \hline
     $Q^b$ & $2\avg{\d\left({K}{C} + f_b\right)\db'}\phantom{\avg{\frac{\p\d}{\p X_i}}}$ \\
     \hline
\end{tabular}
\caption{buoyancy second order structure function}
\end{subtable}
\caption{Summary of terms in the scale-by-scale equations. 
Overbar denotes the two-point mean value (see definitions~\ref{deltaBarOperators}),
while prime and $\mn{.}$ denote fluctuation and mean in a Reynolds averaging
sense.}
\label{termsTab}
\end{table}

\FloatBarrier
\section{Input Data}
\label{secInputData}
Before evaluating the scale-by-scale energy budgets, this section details the 
simulation setup, observational context, and diagnostic statistics. We first 
describe the LES configuration alongside field campaign 
observations. We then outline the data sampling procedures, reference state 
calculations, and single-point baseline profiles used to interpret scale-space 
dynamics.

\subsection{BOMEX and EUREC$\protect^{4}$A}\label{secBOMEX_EURECA}
	We ran a state-of-the-art LES of the BOMEX field campaign (Barbados
	Oceanographic and Meteorological Experiment), cf.~\cite{bomexPaper} with the MicroHH
	code \citep{Hee}. 
    We compare our finding with the EUREC$^{4}$A field campaign (Elucidating the Role of
	Clouds-Circulation Coupling in Climate, see~\cite{Bony2017}), which also took place 
    near Barbados.

The configuration of the BOMEX LES experiment is designed to represent a
specific case of shallow trade wind cumulus convection, which was observed on
22-23 June 1969 during the BOMEX campaign. This was selected because of
remarkably steady behavior of the ABL and no apparent influence of precipitation
or mesoscale circulations~\citep{Siebesma2003}. To retain such a steadiness in the simulation,
special care has been taken to assure that the surface fluxes are balanced by
the vertically integrated prescribed large-scale forcings and radiative cooling. 
The BOMEX simulation was performed at an isotropic grid resolution of 
$10\,\mathrm{m}$ ($10\,\mathrm{m} \times 10\,\mathrm{m} \times 10\,\mathrm{m}$) 
across a domain of $5120\,\mathrm{m} \times 5120\,\mathrm{m} \times 3200\,\mathrm{m}$, 
with periodic boundary conditions applied in the horizontal directions. 
Anelastic equations were solved using a second-order Smagorinsky--Lilly 
subgrid-scale turbulence model. All remaining simulation parameters follow 
\cite{Siebesma2003} (see \cite{Hee} for implementation details). A selection of 
parameters chosen for comparison with airborne measurement conditions is presented 
in Table~\ref{TableIF}.

Clouds in Barbados are considered well-representative
of clouds in trade wind regions in tropical oceans~\citep{Medeiros2016}. 
Consequently, the
comparison with the EUREC$^{4}$A, 
providing higher-resolution measurements
than the BOMEX,
strengthens reliability of the analysis.

The in situ aircraft measurements, previously analyzed in \cite{Nowak2025}, have
been performed over almost two weeks during the EUREC4A campaign (2-13 February
2020). Throughout the period, the regions encountered a range of conditions -
precipitating and non-precipitating clouds, different mesoscale cloud patterns,
mesoscale subsidence or ascent of various strength
\citep{Bony2022,George2021,George2023}. Nevertheless, the lower atmosphere
exhibited consistent vertical structure characteristic of shallow cumulus
convection, i.e. almost well-mixed subcloud layer, conditionally unstable cloud
layer and stable trade-wind inversion. In particular, despite varying cloud
structure inside the cloud layer, the variability of thermodynamic conditions in
the subcloud layer was arguably rather limited \citep{Albright2022}. 

Therefore, the meteorological conditions represented in the simulation fit into
a broader range encountered during EUREC$^4$A campaign. Comparison betweeen BOMEX
parameters and statistics derived from the measurements of the entire EUREC$^4$A
campaign \citep{Albright2022,Bony2022} is presented in Table \ref{TableIF}.
Therein, $u_* = \langle \avg{u'w'}_0^2+\avg{v'w'}_0^2\rangle^{1/4}$ denotes the friction velocity. 
For EUREC$^4$A data this friction velocity was estimated with
formula $u_*=\sqrt{C_D} U_0$, where $C_D = 0.0011$ and
$U_0$ is the mean wind close to the surface, as reported in \cite{Bony2022}.

\begin{table}[h!]
    \centering
    \renewcommand{\arraystretch}{1.25} %
    \begin{tabular}{||l|l|c|r@{ $\pm$ }l||}
        \hline
        \multicolumn{1}{||c|}{Parameter} & \multicolumn{1}{c|}{Units} & \multicolumn{1}{c|}{LES} & \multicolumn{2}{c||}{EUREC$^4$A} \\
        \hline\hline
        $u_*$ & $\mathrm{m\,s^{-1}}$ & $0.28$ & $0.28$ & $0.07$ \\
        $\langle w'\theta'\rangle_0$ & $10^{-3}\,\mathrm{K\,m\,s^{-1}}$ & $8.0$ & $5.5$ & $3.2$ \\
        $\langle w'q'\rangle_0$ & $10^{-5}\,\mathrm{m\,s^{-1}}$ & $5.2$ & $5.75$ & $1.60$ \\
        \hline
    \end{tabular}
    \caption{Comparison of LES boundary conditions \citet{Siebesma2003} with observed conditions during EUREC$^4$A \citep{Albright2022,Bony2022}. Observational values are presented as mean $\pm$ standard deviation.}
    \label{TableIF}
\end{table}
\FloatBarrier
\subsection{LES data samples}
\label{inputSec}
\begin{figure}[h!]
    \centering
    \includegraphics[width=0.6\linewidth]{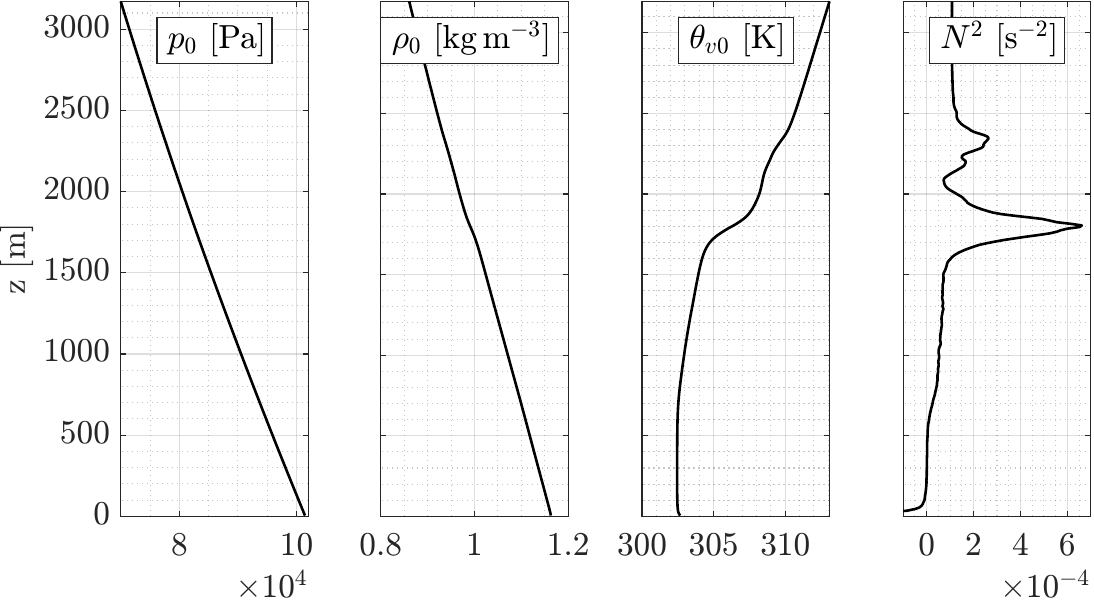}
    \caption{Reference profiles in a main run}
    \label{refProfiles}
\end{figure}

Data samples of physical fields including $\bo{u}$, $b$, $p$, $\nu_s$
were collected from the LES simulation.
A preliminary case was run until reaching a near-stationary level of turbulent kinetic energy (TKE).
At that point, the anelastic reference mean profiles 
in Fig. \ref{refProfiles} were computed by horizontal averaging
and used to initialize the main run. 
This approach allows for minimizing perturbations of the reference profiles and considering 
nearly-stationary TKE at the same time.

\begin{figure}[h!]  
    \centering
    \begin{subfigure}[t]{0.45\textwidth}
        \includegraphics[width=\linewidth]{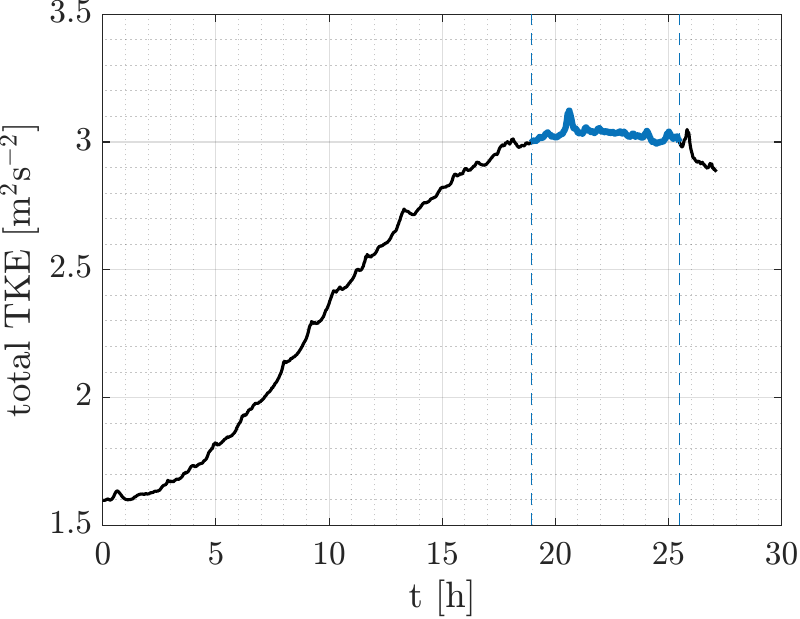}
        \caption{TKE evolving in time.
        Time interval under consideration marked blue.}
        \label{tkeFig}
    \end{subfigure}
    \hfill
    \begin{subfigure}[t]{0.46\textwidth}
        \includegraphics[width=\linewidth]{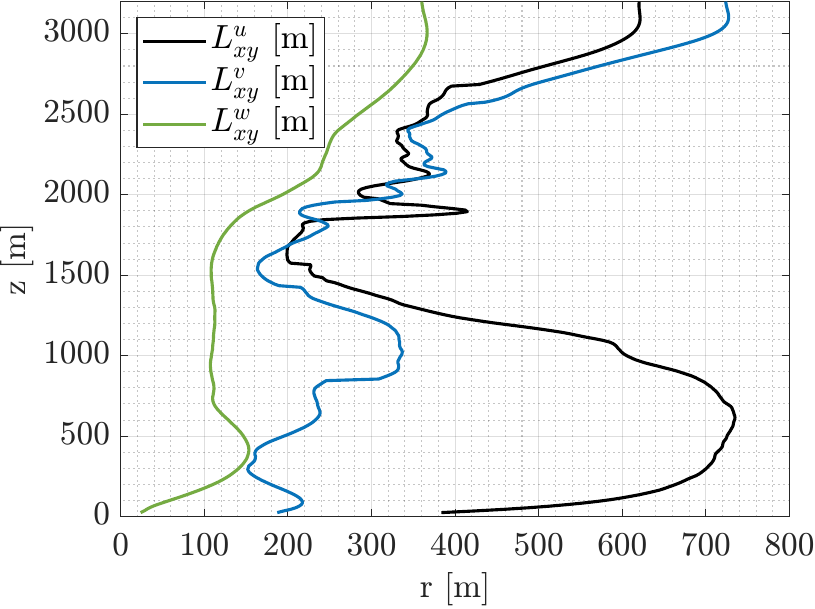}
        \caption{Horizontal integral length scales, expressed as a characteristic 
        correlation radius $r$.}%
        \label{LFig}
    \end{subfigure}
    \vspace{0.5cm}
    \begin{subfigure}[t]{0.46\textwidth}
        \includegraphics[width=\linewidth]{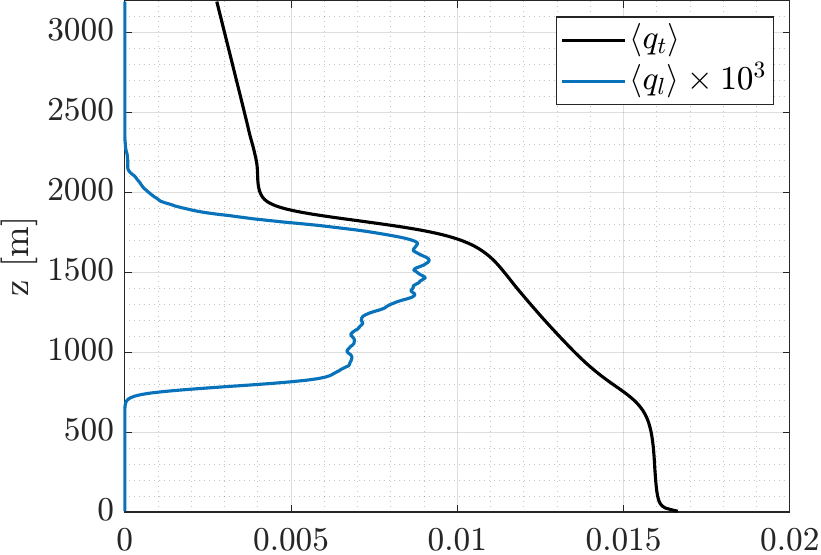}
    \caption{Liquid water mixing ratio averaged and total water mixing ratio}
    \label{qlMeanFig}
    \end{subfigure}
    \hfill
    \begin{subfigure}[t]{0.45\textwidth}
        \includegraphics[width=\linewidth]{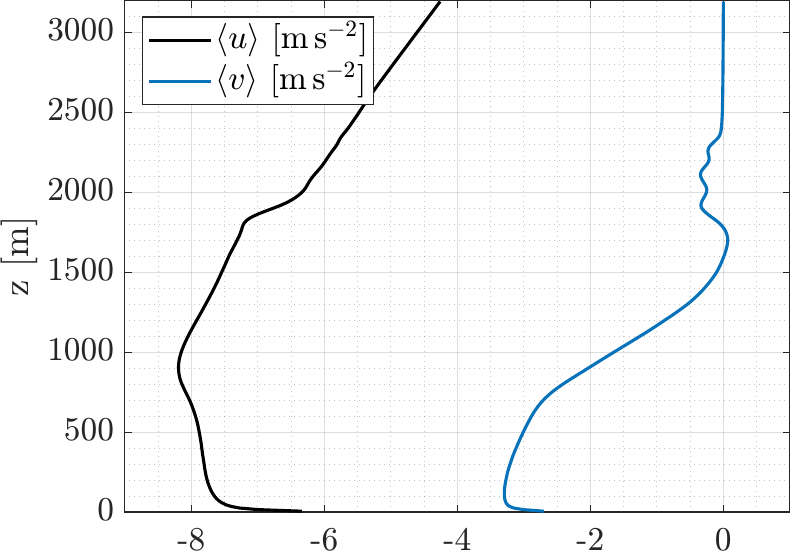}
    \caption{Mean velocity profiles. Vertical component not shown as equal zero.}
    \label{uvMeanFig}
    \end{subfigure}
    \caption{Selected characteristics of the input data.}
\end{figure}

The snapshots were sampled from the period of near-stationary level of TKE (between
$t = 18$h and $t=25$h), marked blue in Fig. (\ref{tkeFig}), with a separation time of 600s between
samples (skipping the first sample). This value is roughly equal to the large-eddy turnover time.
The time derivatives are computed with backward difference at the first substep of the Runge-Kutta scheme 
used by MicroHH. This gives the exact instantaneous time derivatives at each (complete) step. In the case of the 
velocity equation, the same result could be obtained by summing all the other terms in
equation \ref{uSummary}. However, the former 
approach allows checking if the budget is closed. The buoyancy equation, on the other hand, gives no such 
possibility because it is not solved by MicroHH directly. We extract the buoyancy from the $w$ equation, 
compute its time derivative as described above, recover the other terms and assign the imbalance to the source term.

To establish the bulk characteristics of the flow, several mean profiles are 
computed. The vertical distributions of liquid water mixing ratio $q_l$ (see Appendix \ref{Appendix}), 
total water mixing ratio $q_t = q_v + q_l$ %
and mean velocities are shown in 
Figures~\ref{qlMeanFig} and \ref{uvMeanFig}. 
\FloatBarrier

To provide a characteristic reference scale for interpreting the scale-by-scale 
budgets, we compute integral length scales from spatial velocity 
autocorrelation functions. Due to strong anisotropy and inhomogeneity of the 
convective boundary layer, the correlation ranges cannot be easily reduced to a 
single representative value. We define the total separation distance between two 
points as $2r$, where $r$ represents the characteristic correlation radius. We 
then compute the directional integral length scales as:

\begin{equation*}
    L_{x}^u (z) = \int_0^{r_0} \frac{\avg{ u'(x, y, z, t) \, u'(x + r, y, z, t)}}{ {u'(x,y,z,t)^2}} \, dr
\end{equation*}
and $r_0$ is the first zero-crossing point where the integrand vanishes. 
The components $L_{y}^u$, $L_{x}^v$, $L_{y}^v$, $L_{x}^w$, and $L_{y}^w$ are 
computed analogously. To simplify the directional variations, these components 
are combined into horizontal scales:
\begin{equation*}
    L_{xy}^u = \sqrt{{L_x^u}^2 + {L_y^u}^2}, \qquad
    L_{xy}^v = \sqrt{{L_x^v}^2 + {L_y^v}^2}, \qquad
    L_{xy}^w = \sqrt{{L_x^w}^2 + {L_y^w}^2},
\end{equation*}
which are plotted in Fig.~\ref{LFig}. These integral length scales serve as a 
measure of the spatial correlation distance, defining the upper scale boundaries 
below which coherent turbulent dynamics can be meaningfully analyzed.

\begin{figure}[ht!] 
    \centering
    \begin{subfigure}[b]{0.49\textwidth}
        \includegraphics[width=\linewidth]{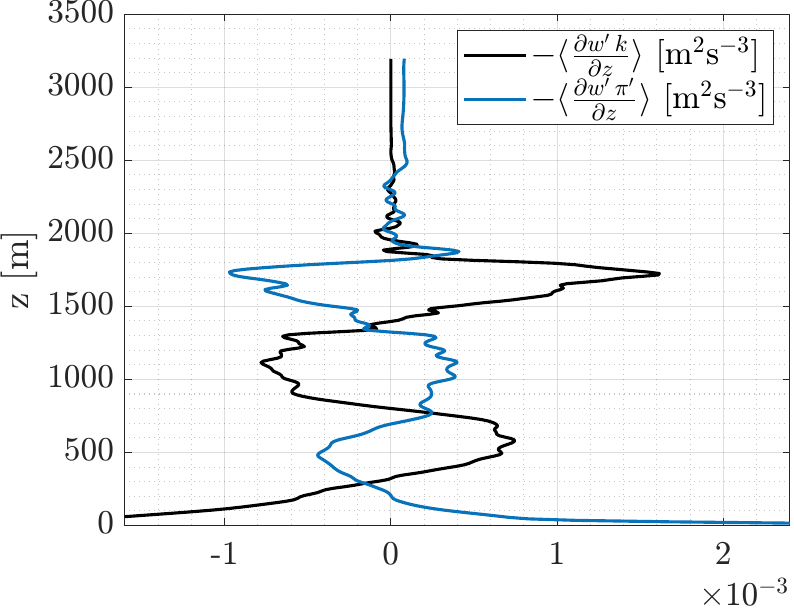}
    \caption{Transport by advection and pressure}
    \label{TuTpMeanFig}
    \end{subfigure}
    \hfill
    \begin{subfigure}[b]{0.49\textwidth}
        \includegraphics[width=\linewidth]{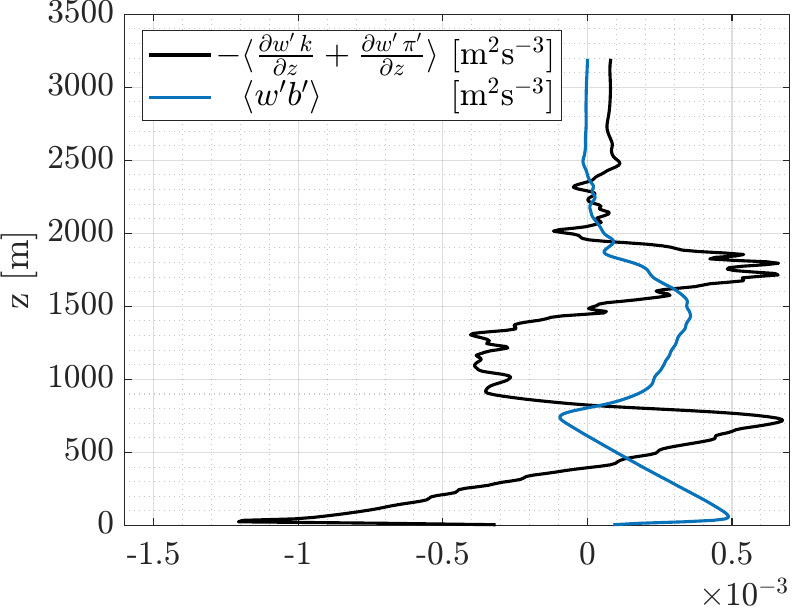}
    \caption{Transport and buoyancy production}
    \label{TuWMeanFig}
    \end{subfigure}
    \caption{Selected terms of the one-point TKE transport equation averaged horizontally and in time.}
    \label{onePointTKEFig}
\end{figure}

Finally, selected one-point statistics related to the TKE transport equation were also computed 
(Fig.~\ref{onePointTKEFig}), as a reference for the scale-by-scale analysis of 
further sections. This includes vertical TKE transport $\avg{\p{w'k}/\p z}=1/2 \avg{\p{w'u'_iu'_i}/\p z}$, pressure transport 
$\avg{\p w' \pi/\p z}$ and buoyancy work $\avg{w' b'}$.

\newcommand{\cmSize}{0.45} %
\FloatBarrier
\section{Results}
\label{secResults}
We begin our discussion by validating the present results against the
one-dimensional field campaign measurements in
Sec.~\ref{secNowak}. Following this comparison, we present a
comprehensive overview of the full three-dimensional budget terms from
LES for both the velocity $\du'^2$ and
buoyancy $\db'^2$ fields in Sec.~\ref{secColormaps}. We then isolate
and analyze selected features of the scale-space
dynamics before concluding with phenomenological discussion
of the plausible mechanisms driving the TKE
and TPE budgets in Sec.~\ref{secLogPlots}.

\FloatBarrier
\subsection{Comparison with the field campaign}
\label{secNowak}
We first validate our data by reference to the study
of~\cite{Nowak2025}, which is based on airborne measurements from the
EUREC$^4$A campaign. This field campaign dataset has different limitations
compared to the current LES results. In EUREC$^4$A, data samples are
available exclusively along 1D tracks at four distinct heights
relative to the boundary layer (BL) depth. For this reason, the
analysis by~\cite{Nowak2025} relied on the assumption of local
isotropy and crude estimates of the derivatives for $T_u$. On the
other hand, the LES approach undertaken here, suffers from modelling
uncertainties.

Apart from these inherent limitations, the sampling methodology used in
EUREC$^4$A and the LES are also different. The LES domain for BOMEX is
relatively small (5~$\times$~5 km~$\times$~3.2 km) but offers vast amount of data along all
directions. In contrast, during EUREC$^4$A, the aircraft penetrated through an
extended spatial domain \citep[about 120~$\times$~20~km;][]{Bony2022}. Hence, the
measurement segments were relatively distant in both space and time. While this
provides good statistical independence, it is likely that the atmospheric
conditions were not entirely homogeneous even along a given flight segment.

Despite these differences, a meaningful comparison of the scale-by-scale budgets
can be performed as long as we consider the same relative heights. These heights
are defined according to the vertical structure of the ABL, taking into account
the contrast in atmospheric conditions and sampling methodologies. Specifically,
we distinguish four levels: near-surface, mid-subcloud, top-subcloud, and cloud
base.

Following the flight pattern, the selected LES levels are chosen to
roughly match the experimental setup:
\begin{itemize}
    \item near-surface: 60~m,
    \item mid-subcloud: two-thirds of the mixed layer top height,
    \item top-subcloud: midway between the mixed layer top and the cloud base,
    \item cloud base: the level of maximum cloud fraction.
\end{itemize}

The mixed layer top was diagnosed using the method from \cite{Albright2022} 
applied to the potential temperature profile. In the quasi-stationary part of 
the LES considered here (from $t=18$~h to $25$~h), this height varied from 
$575$~m at $t=18$~h to $635$~m at $t=25$~h. This is comparable to the mean mixed 
layer top of $555$~m observed during EUREC$^4$A, as shown in 
Table~\ref{Table_comp}.

The cloud base during EUREC$^4$A, as considered in \cite{Nowak2025}, was located 
at $807$~m above the ground, which is consistent with Fig.~\ref{qlMeanFig}. In 
the LES, the cloud base height increased slightly over time, rising from 
$775$~m at $t=18$~h to $825$~m at $t=25$~h. For subsequent comparisons, we use 
the mean value of approximately $800$~m. The estimates for the Lifting 
Condensation Level (LCL), top-subcloud and mid-subcloud layers, as well as the 
inversion layer top, are also included in Table~\ref{Table_comp}.

\begin{table}[h!]
    \centering
    \renewcommand{\arraystretch}{1.25} %
    \begin{tabular}{||l|c|r@{ \text{---} }l|r@{ $\pm$ }l||}
        \hline
        \multicolumn{1}{||c|}{Parameter} & \multicolumn{1}{c|}{Units} & \multicolumn{2}{c|}{LES ($t$: $18\text{---}25\,\mathrm{h}$)} & \multicolumn{2}{c||}{EUREC$^4$A} \\
        \hline\hline
        mixed layer top      & m & \multicolumn{2}{c|}{$575\text{---}635$} & 555 & 79 \\
        cloud base           & m & \multicolumn{2}{c|}{$775\text{---}825$} & 807 & 81 \\
        LCL                  & m & \multicolumn{2}{c|}{$749\text{---}803$} & 694 & 105 \\
        top subcloud layer   & m & \multicolumn{2}{c|}{$675\text{---}730$} & 708 & 84 \\
        middle cloud layer   & m & \multicolumn{2}{c|}{$383\text{---}423$} & 370 & 52 \\
        near-surface layer   & m & \multicolumn{2}{c|}{$60\text{---}60$}  & 60  & \text{---} \\
        inversion layer top  & m & \multicolumn{2}{c|}{$2000\text{---}2000$} & 2762 & 432 \\
        \hline
    \end{tabular}
    \caption{Comparison of mixed layer, liquid condensation level, 
    subcloud, and inversion layer heights between LES and observed conditions during EUREC$^4$A \citep{Albright2022,Bony2022}. 
    For LES, ranges correspond to values spanning $t = 18\,\mathrm{h}$ to $25\,\mathrm{h}$. 
    Observational values are presented as mean $\pm$ standard deviation.}
    \label{Table_comp}
\end{table}

It can be concluded that the atmospheric conditions in the BOMEX LES and the
EUREC$^4$A campaign are comparable within uncertainty bounds. However, the
horizontal flight levels considered in~\cite{Nowak2025}, referred to as the
mid-subcloud and top-subcloud levels, were performed at somewhat lower
altitudes ($292$~m and $595$~m, respectively) than the estimated values
presented in Table~\ref{Table_comp}. Hence, to ensure a meaningful comparison,
we use roughly the same absolute altitudes as in~\cite{Nowak2025}.
Furthermore, for this comparison, we adopt the constrained methodology of
\cite{Nowak2025}, where statistics are calculated along 1D trajectories. In
the LES, this is replicated by sampling data along 1D horizontal grid lines
across the computational domain. We treat this qualitative comparison as a
measure of consistency between both datasets.

For each height, the terms $W$ and $S_3$ (defined exactly as in \cite{Nowak2025}, 
therefore not as $W^u$ and $S_3^u$ in \ref{termsTab}) 
are computed along all lines in the
$x$-direction and all lines in the $y$-direction, after which the spatial
average is taken. The transport term $T_u$ is calculated in the same manner,
and its vertical gradients are estimated using these four discrete levels
exclusively to match the reference methodology. The only methodological
difference lies in the estimation of the dissipation term. In this study,
dissipation is determined via the transfer to subgrid-scales computed using
local 3D derivatives, which were unavailable in the airborne dataset.

\begin{figure}[htbp]
     \centering
    \begin{subfigure}[b]{0.45\textwidth}
         \centering
         \includegraphics[width=\textwidth,trim={0, 6mm, 0, 0}, clip]
         {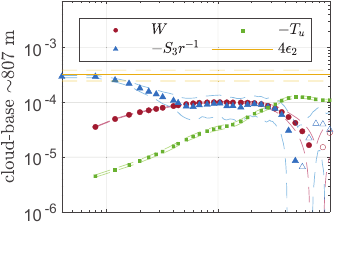}
     \end{subfigure}
     \hfill
     \begin{subfigure}[b]{0.42\textwidth}
         \centering
         \includegraphics[width=\textwidth,trim={0, -3mm, -3mm, 0}, clip]
         {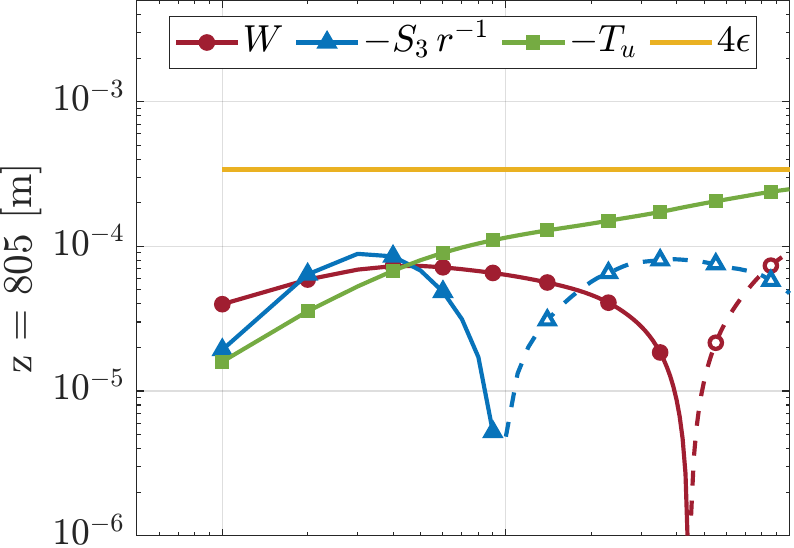}
     \end{subfigure}
    \begin{subfigure}[b]{0.45\textwidth}
         \centering
         \includegraphics[width=\textwidth,trim={0, 6mm, 0, 0}, clip]
         {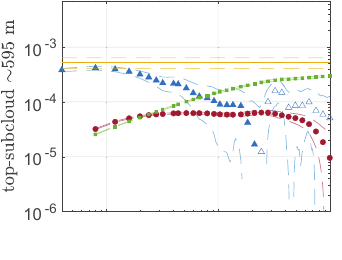}
     \end{subfigure}
     \hfill
     \begin{subfigure}[b]{0.42\textwidth}
         \centering
         \includegraphics[width=\textwidth,trim={0, -3mm, -3mm, 0}, clip]
         {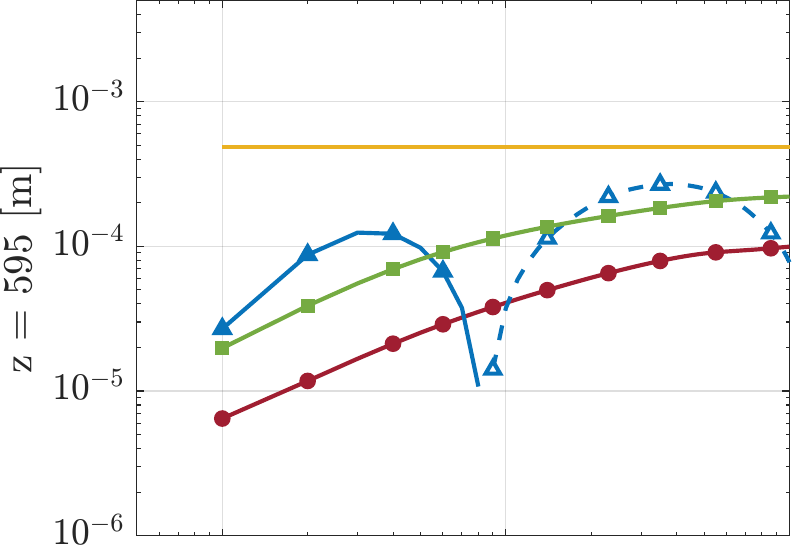}
     \end{subfigure}
    \begin{subfigure}[b]{0.45\textwidth}
         \centering
         \includegraphics[width=\textwidth,trim={0, 6mm, 0, 0}, clip]
         {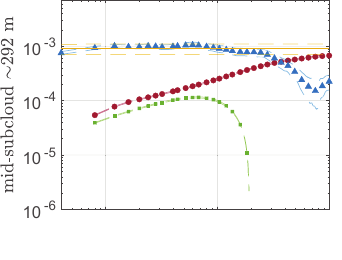}
     \end{subfigure}
     \hfill
     \begin{subfigure}[b]{0.42\textwidth}
         \centering
         \includegraphics[width=\textwidth,trim={0, -3mm, -4mm, 0}, clip]
         {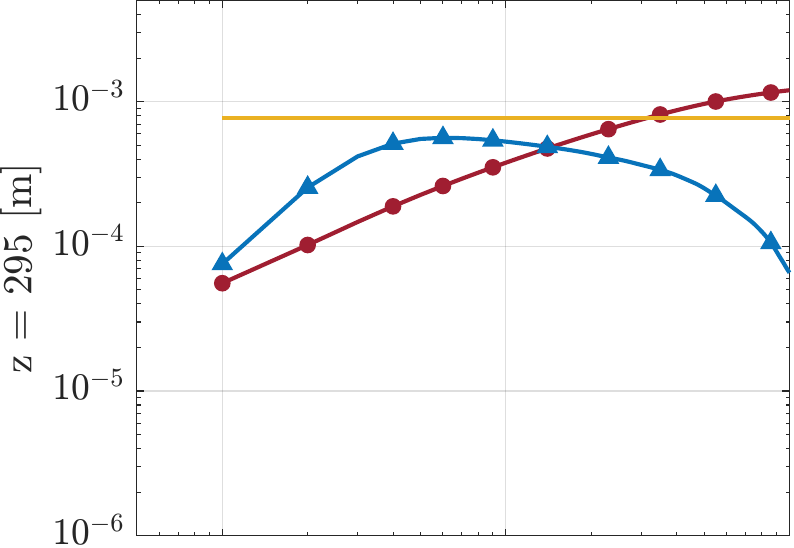}
     \end{subfigure}
    \begin{subfigure}[b]{0.45\textwidth}
         \centering
         \includegraphics[width=\textwidth,trim={0, 0, 0, 0}, clip]
         {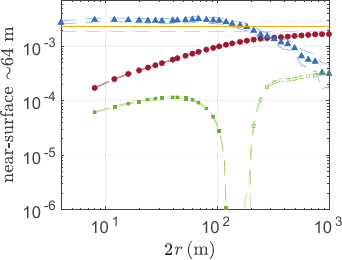}
     \end{subfigure}
     \hfill
     \begin{subfigure}[b]{0.42\textwidth}
         \centering
         \includegraphics[width=\textwidth,trim={0, -3mm, 0, 0}, clip]
         {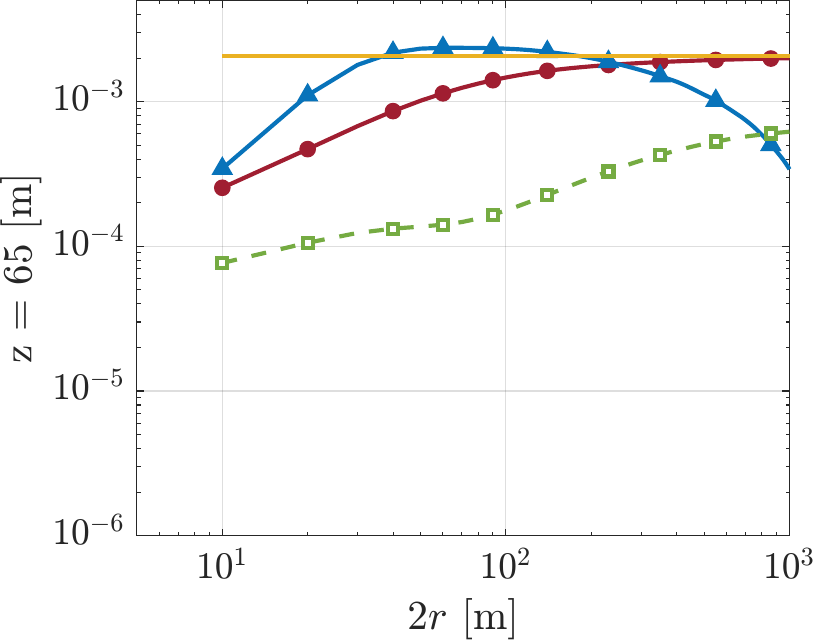}
     \end{subfigure}
     \caption{
     Selected terms of the kinetic energy budget appearing in \cite{Nowak2025}, but calculated
     with our LES data (right column), 
     compared with the field campaign results of \cite{Nowak2025} (left column).
     Negative values denoted with dashed lines and white markers.
     Note that the notation for the separation distance in the left column was changed 
     (with respect to the original paper) to match the convention in the current study.}
     \label{nowakComparisonFig}
\end{figure}

The comparison of the results is presented in Fig.~\ref{nowakComparisonFig},
showing a satisfactory level of consistency. The dissipation rate decreases
with height from the near-surface level to the cloud base, which is in general
agreement with the experimental measurements.

Regarding the $S_3 r^{-1}$ term, its sudden drop at small scales in all right-hand
plots is an expected artifact of the Smagorinsky subgrid-scale model. Aside from
this numerical drop, the inter-scale transport looks very similar at $z
\approx 65$~m and $z \approx 295$~m. At greater heights, both studies reveal a
change of sign in this term; however, the LES predicts this transition at
smaller scales than the observations.

The buoyancy production term $W$ also shows reasonable agreement across most
heights. Some discrepancy can be observed at $z \approx 595$~m, which can be
partially attributed to a slight height misalignment between the datasets. In
the left column (measurements), $W$ has already begun to drop at large scales, similar
to its behavior at $z \approx 807$~m. Conversely, in the right column (LES), it
still more closely resembles the profile at $z \approx 295$~m. In the cloud-base
region, both datasets predict a change of sign occurring at a similar scale
size.

The inter-space transport term $T_u$ appears rather inconsistent at lower heights
($z \approx 65$~m and $z \approx 295$~m), though this should be interpreted
with caution. The estimation of the vertical derivative relies solely on four
discretized height levels, which likely amplifies any underlying discrepancies.
Furthermore, this term is highly sensitive to the choice of numerical
differentiation (e.g., using spline interpolation resulted in noticeably
different characteristics). Because the sign of $T_u$ changes between $z \approx
65$~m and $z \approx 595$~m, the linear differentiation method yields a near-zero 
value at $z \approx 295$~m. For this reason, the green line is absent at this
level in the right column and largely missing in the left column. At higher
altitudes, the agreement of $T_u$ improves, particularly at $z \approx 595$~m.

We conclude this section by noting that despite these discrepancies, the
LES dataset demonstrates sufficient reliability to support further qualitative 
analysis of the scale-by-scale energy budget.

\FloatBarrier
\subsection{Dimensionality reduction and overview of new results}
\label{secColormaps}

Proceeding to a more detailed analysis of the LES data, we face the challenge 
of visualizing the results within the full, four-dimensional space of 
${r_x, r_y, r_z, Z}$.

To handle the $r$-dimensions in a way that is conceptually simple yet physically 
meaningful, we focus on the net inter-scale budget over isotropic scales. 
Although we do not expect the turbulence to be fully isotropic, retaining this 
direction-independent notion of scale allows for a straightforward comparison 
with theory.

The analysis is conducted by averaging each term $\phi$ from Table~\ref{termsTab} 
within spherical volumes:
\begin{equation}
    \phi(Z, r) = \frac{3}{4\pi r^3}\int_{\Omega(r)} \phi(Z, r_x, r_y, r_z) dV,
\end{equation}
\begin{equation*}
    \Omega(r): r_x^2 + r_y^2 + r_z^2 < r^2.
\end{equation*}
This represents an exact evaluation of the approach proposed in \cite{Nowak2025}, 
where it could only be roughly estimated due to the data limitations of that study.

\begin{figure}[h!]
     \centering
    \begin{subfigure}[b]{0.45\textwidth}
         \centering
         \includegraphics[width=\textwidth]{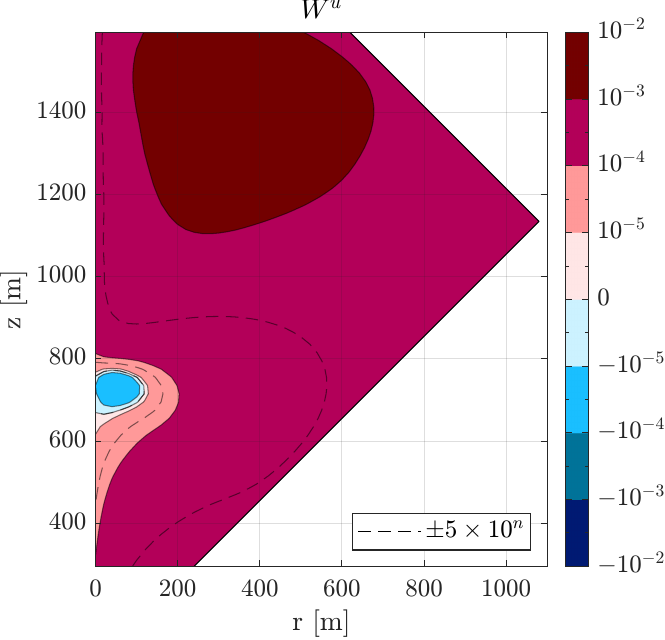}
     \end{subfigure}
     \hfill
     \begin{subfigure}[b]{0.45\textwidth}
         \centering
         \includegraphics[width=\textwidth]{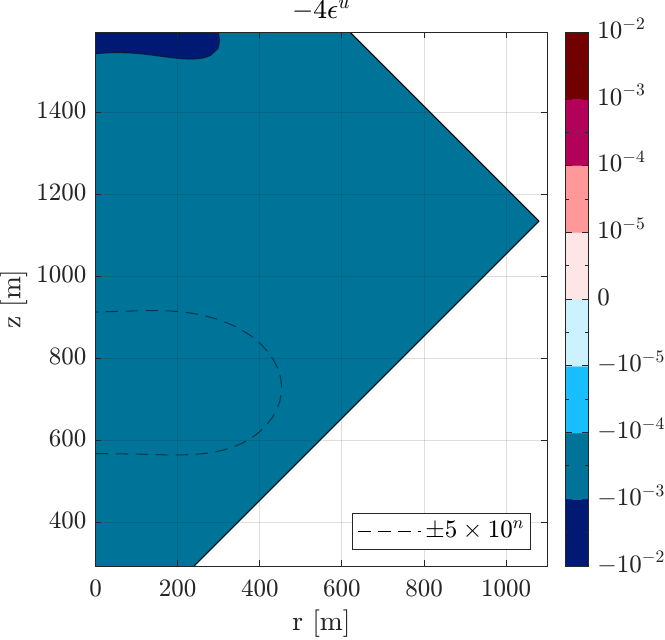}
     \end{subfigure}
     \caption{Spherical-averaged dominant terms in the $\du'^2$ equation.}
     \label{uColormapsWE}
\end{figure}

\begin{figure}[h!]
     \begin{subfigure}[b]{0.45\textwidth}
         \centering
         \includegraphics[width=\textwidth]{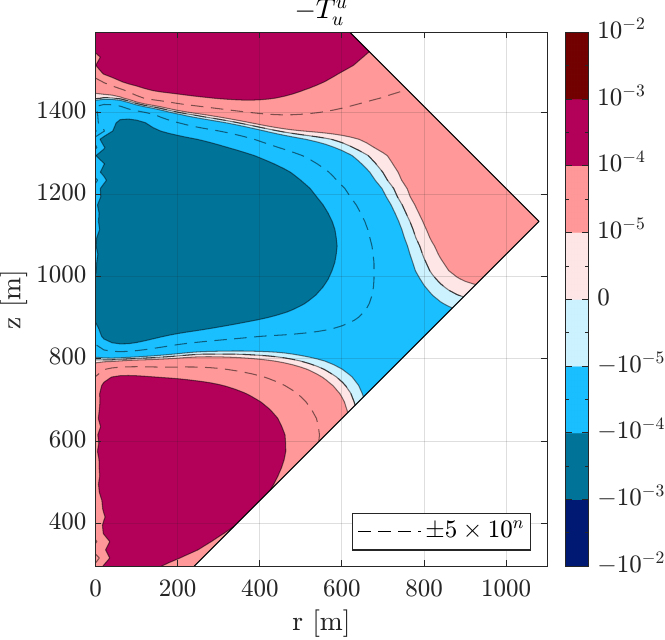}
     \end{subfigure}
     \hfill
     \begin{subfigure}[b]{0.45\textwidth}
         \centering
         \includegraphics[width=\textwidth]{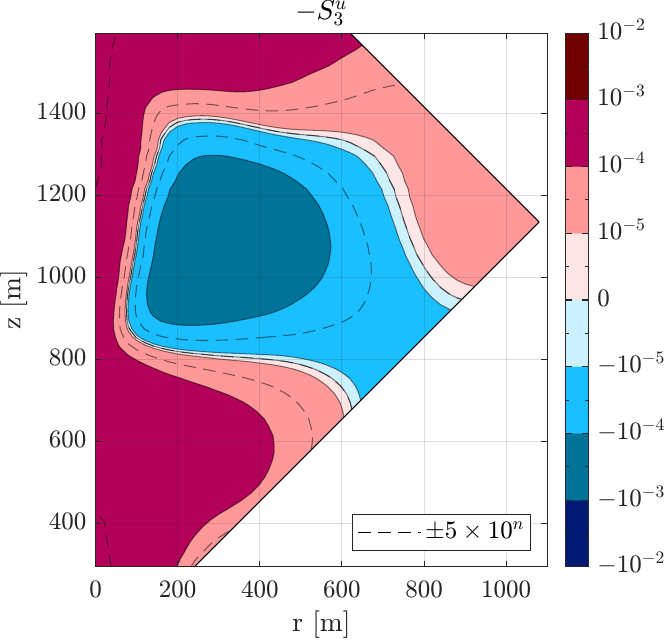}
     \end{subfigure}
    \begin{subfigure}[b]{0.45\textwidth}
         \centering
         \includegraphics[width=\textwidth]{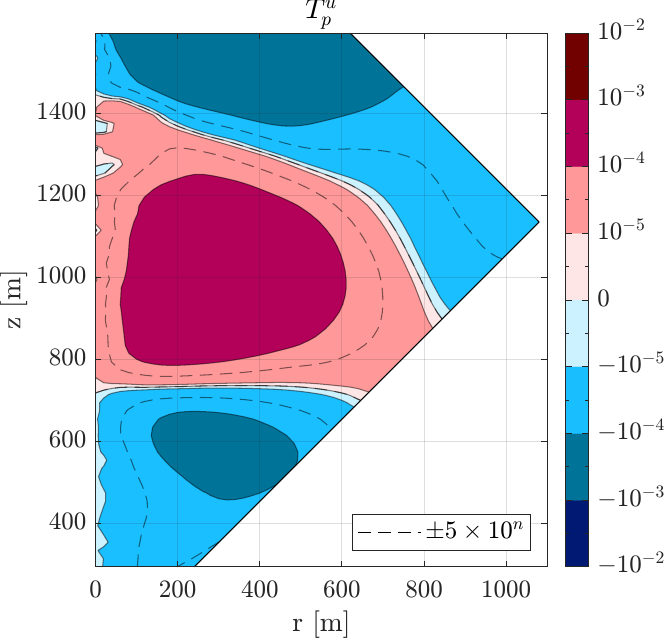}
     \end{subfigure}
     \hfill
     \begin{subfigure}[b]{0.45\textwidth}
         \centering
         \includegraphics[width=\textwidth]{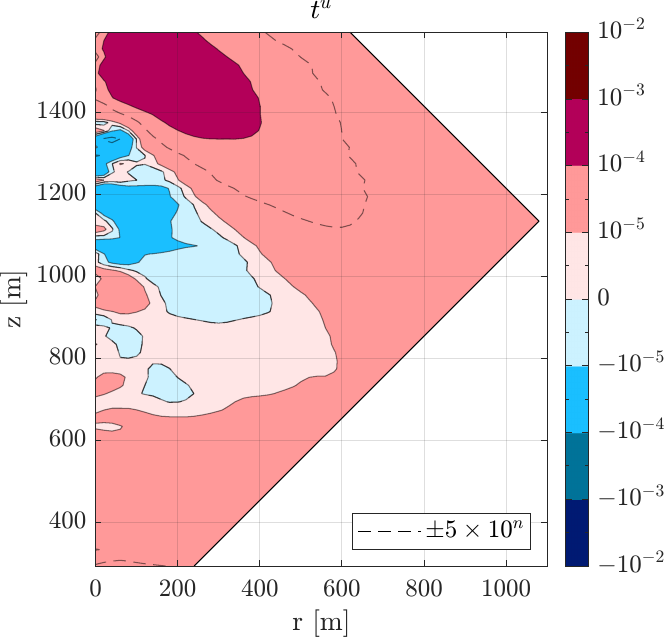}
     \end{subfigure}
    \centering
    \begin{subfigure}[b]{\cmSize\textwidth}
         \centering
         \includegraphics[width=\textwidth]{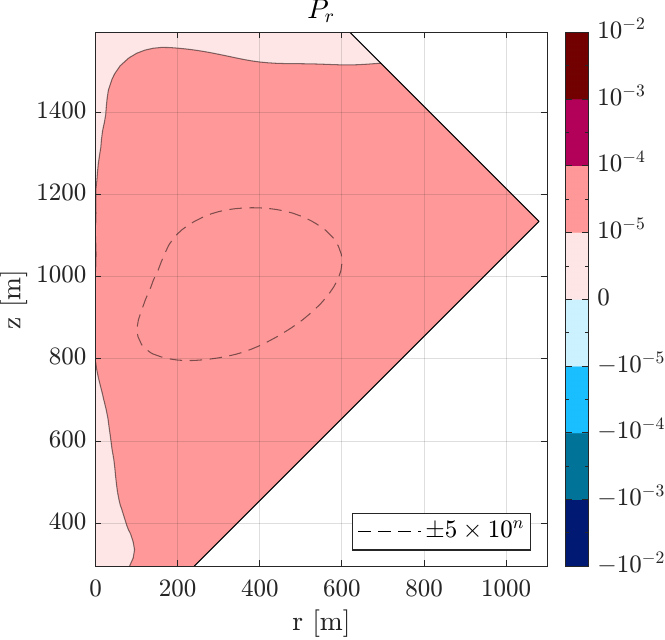}
    \end{subfigure}
    \hfill
    \begin{subfigure}[b]{\cmSize\textwidth}
         \centering
         \includegraphics[width=\textwidth]{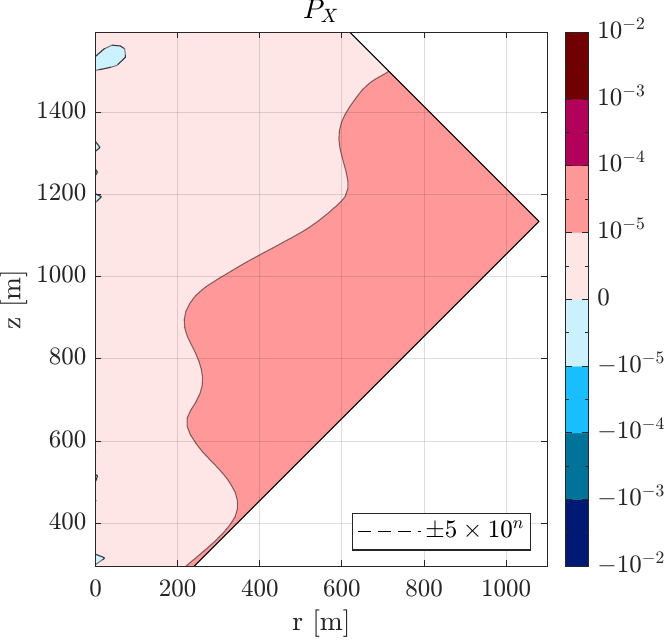}
    \end{subfigure}
        \caption{Spherical-averaged significant terms in the $\du'^2$ budget.}  
        \label{uColormaps1}
\end{figure}

\begin{figure}[h!]
     \centering
    \begin{subfigure}[b]{\cmSize\textwidth}
         \centering
         \includegraphics[width=\textwidth]{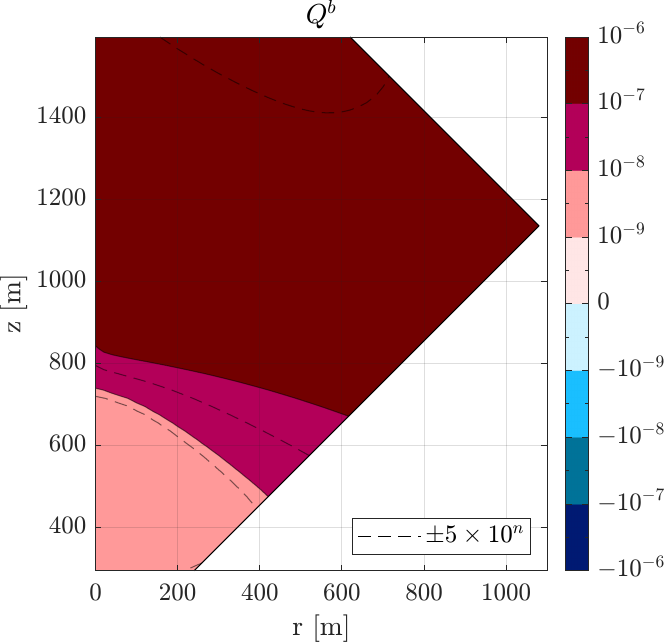}
     \end{subfigure}
     \hfill
     \begin{subfigure}[b]{\cmSize\textwidth}
         \centering
         \includegraphics[width=\textwidth]{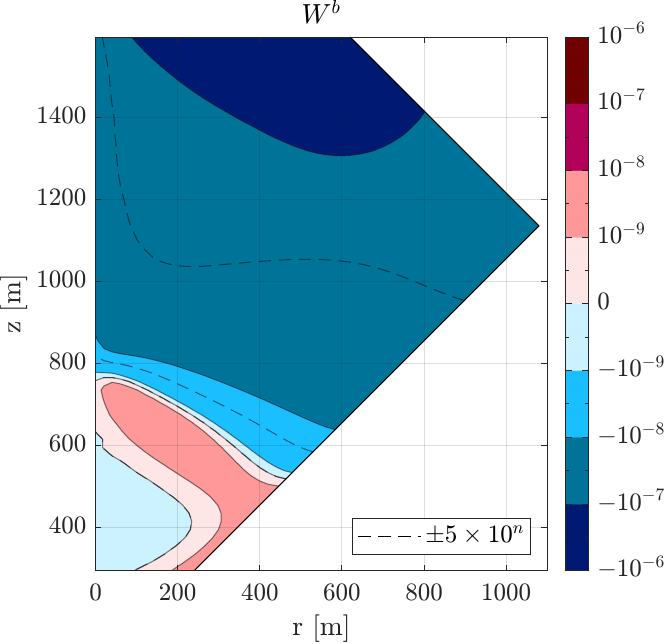}
     \end{subfigure}
     \begin{subfigure}[b]{\cmSize\textwidth}
         \centering
         \includegraphics[width=\textwidth]{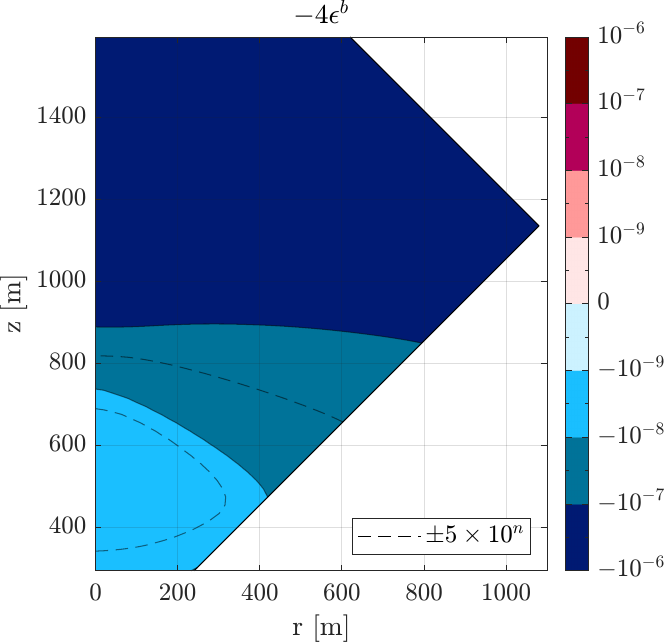}
     \end{subfigure}
     \hfill
     \begin{subfigure}[b]{\cmSize\textwidth}
         \centering
         \includegraphics[width=\textwidth]{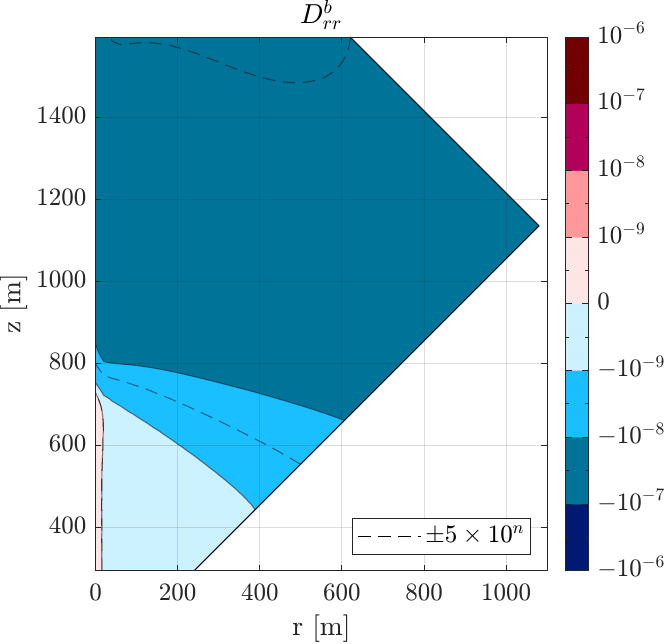}
     \end{subfigure}
        \caption{Spherical-averaged terms in the $\db'^2$ budget.}  
        \label{bColormaps1}
\end{figure}

\begin{figure}[h!]
     \centering
    \begin{subfigure}[b]{\cmSize\textwidth}
         \centering
         \includegraphics[width=\textwidth]{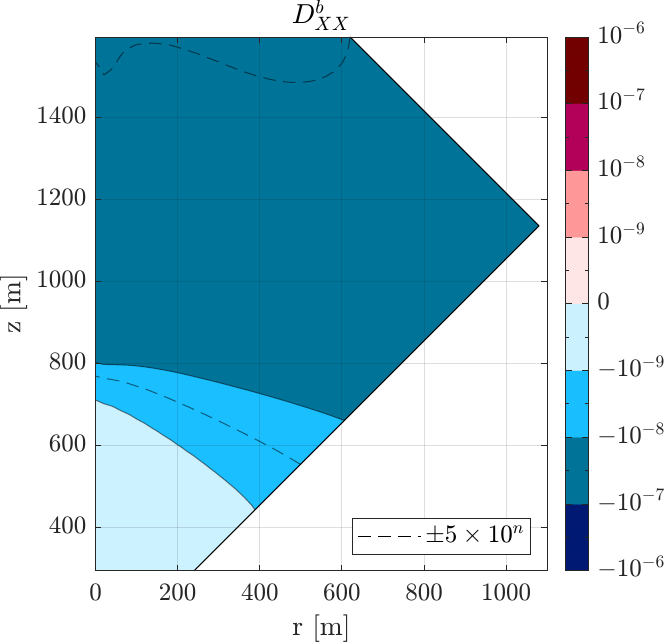}
     \end{subfigure}
     \hfill
     \begin{subfigure}[b]{\cmSize\textwidth}
         \centering
         \includegraphics[width=\textwidth]{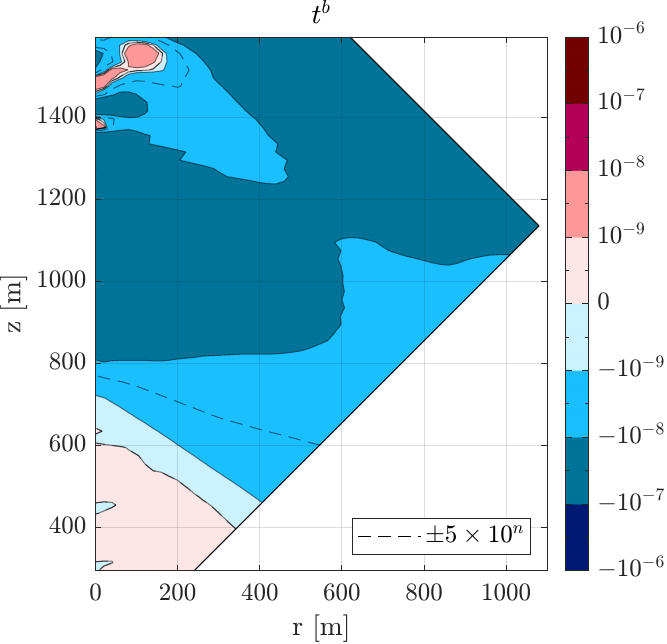}
     \end{subfigure}
     \begin{subfigure}[b]{\cmSize\textwidth}
         \centering
         \includegraphics[width=\textwidth]{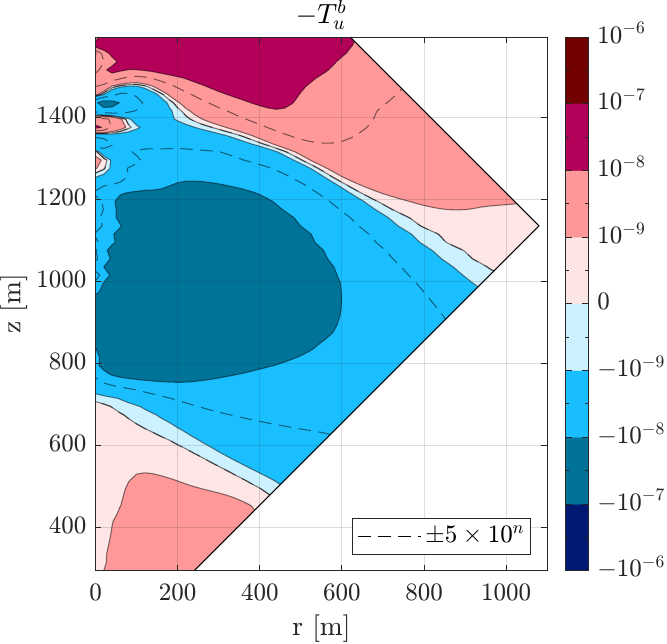}
     \end{subfigure}
     \hfill
     \begin{subfigure}[b]{\cmSize\textwidth}
         \centering
         \includegraphics[width=\textwidth]{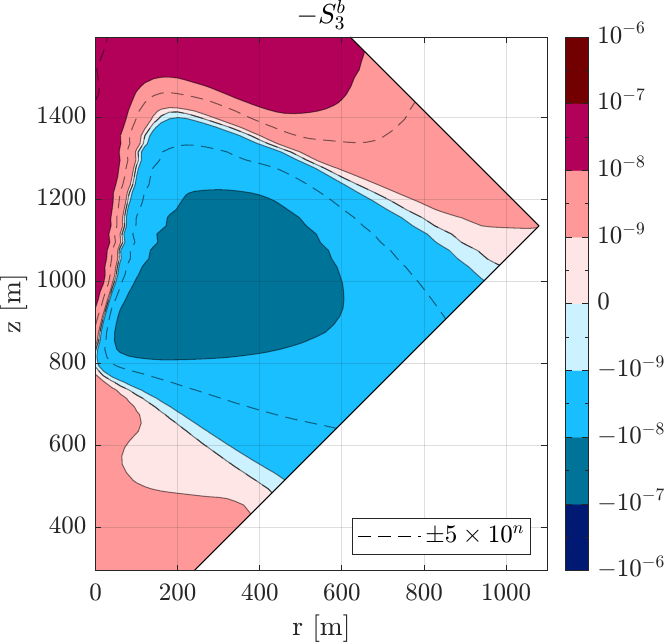}
     \end{subfigure}
    \begin{subfigure}[b]{\cmSize\textwidth}
         \centering
         \includegraphics[width=\textwidth]{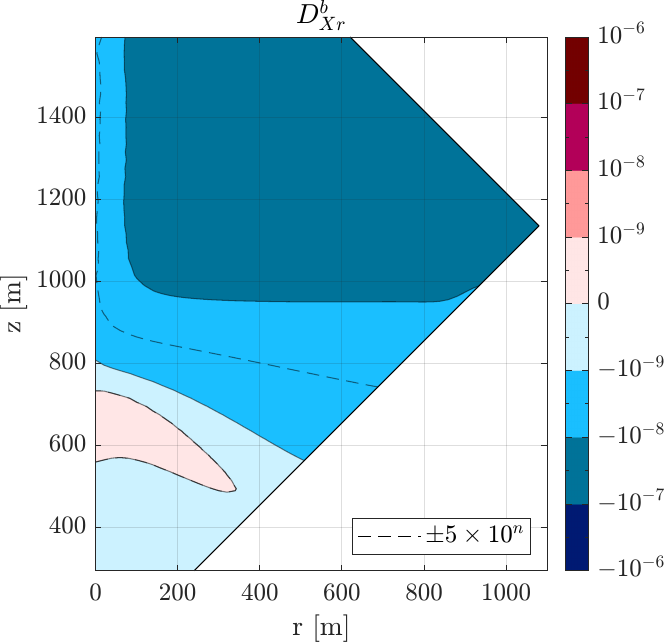}
     \end{subfigure}
     \hfill
     \begin{subfigure}[b]{\cmSize\textwidth}
         \centering
         \includegraphics[width=\textwidth]{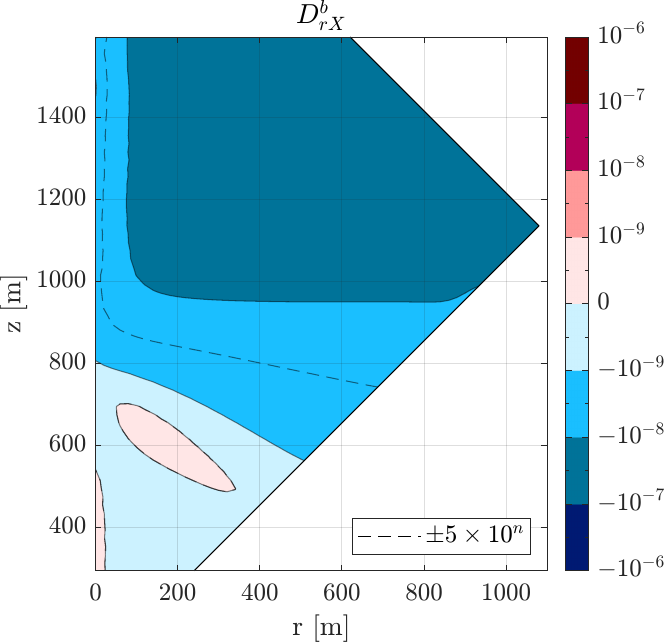}
     \end{subfigure}
        \caption{Spherical-averaged terms in $\db'^2$ budget.} 
        \label{bColormaps2}
\end{figure}

The selected resulting net terms (in SI units) are presented as colormaps in 
Figs.~\ref{uColormapsWE} and \ref{uColormaps1} 
for the velocity structure function 
and in Figs.~\ref{bColormaps1} and \ref{bColormaps2} for buoyancy. We also 
computed the standard errors of independent samples for each field. These 
errors were found to be at least two orders of magnitude lower than the field 
values themselves, and are therefore considered insignificant. Consequently, the 
primary sources of uncertainty in this study are related to modeling rather 
than lack statistical convergence.

Inspecting the results, we observe that the budget of $\du'^2$ is dominated by 
$W^u$, $4\varepsilon^u$, $S_3^u$, $T_u^u$, $T_p^u$, and to some extent the transient 
term $t^u$ (Fig.~\ref{uColormapsWE} and \ref{uColormaps1}). Worth noting is the fact that 
$S_3^u$, $T_u^u$ and $T_p^u$ all change sign in the cloud layer.
Turbulence productions rates $P_r^u$ and $P_X^u$ are also non-negligible but slightly smaller
than the other terms mentioned.
The remaining terms play a secondary role, staying below $10^{-4}$ 
and have therefore been omitted from the figures.

\FloatBarrier

\FloatBarrier
In the case of the $\delta b'^2$ budget (Fig.~\ref{bColormaps1} and \ref{bColormaps2}), 
the most important terms are $Q^b$, $W^b$, 
and $4\varepsilon^b$. The contribution of the subgrid-scale terms $D^b_{rr}$, $D^b_{XX}$,
$D^b_{Xr}$ and $D^b_{rX}$ 
is more significant 
here than in the $\delta u'^2$ budget, surpassing $-S_3^b$ and $-T_u^b$. Because 
these subgrid-scale terms do not represent well-resolved physical phenomena but rather subgrid-scale 
parameterizations, they are interpreted here as a measure of model uncertainty.
Instead, we focus on the directly resolved terms, including 
$-S_3^b$ and $-T_u^b$, particularly because they both change sign within the cloud layer.

The values of $Q^b$ below the lifting condensation level are relatively high 
due to several factors. First, this term accounts for large-scale forcing, including 
contributions from subsidence velocity advection. As described in 
Appendix~\ref{secNumericalMethod}, $Q^b$ is obtained from the buoyancy equation 
residual, meaning it also accumulates numerical errors. Finally, the spherical 
averaging process inherently combines contributions from multiple heights into 
a single net scale. Nevertheless, the dominant driver remains the phase changes 
above the LCL ($\approx 800$~m), which exceed all other contributions by two 
orders of magnitude.

Following this overview %
we proceed to a more in-depth investigation of its specific features.


\FloatBarrier
\subsection{Inhomogeneity}

\begin{figure}[htbp]
     \centering
     \begin{subfigure}[b]{\cmSize\textwidth}
             \centering
             \includegraphics[width=\textwidth]
             {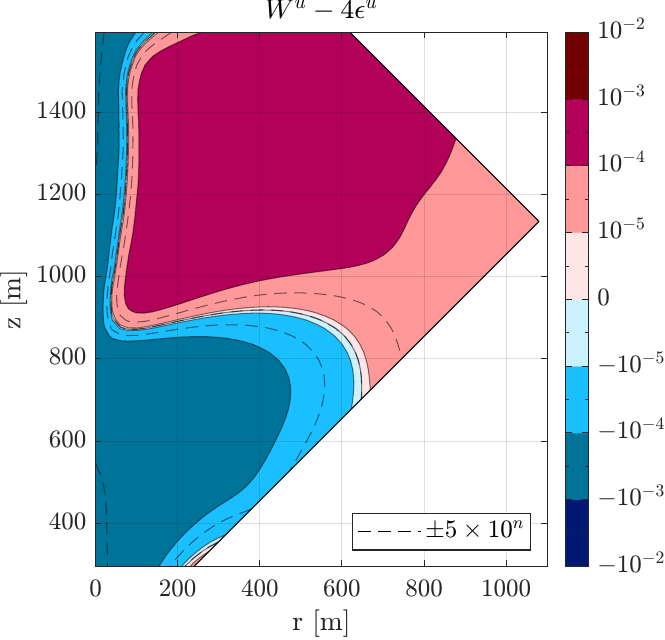}
             \caption{Local energy surplus fueling the interspace transfer}
             \label{WorkDissipDiffFig}
          \end{subfigure}
          \hfill
    \begin{subfigure}[b]{\cmSize\textwidth}
        \centering
        \includegraphics[width=\textwidth]
        {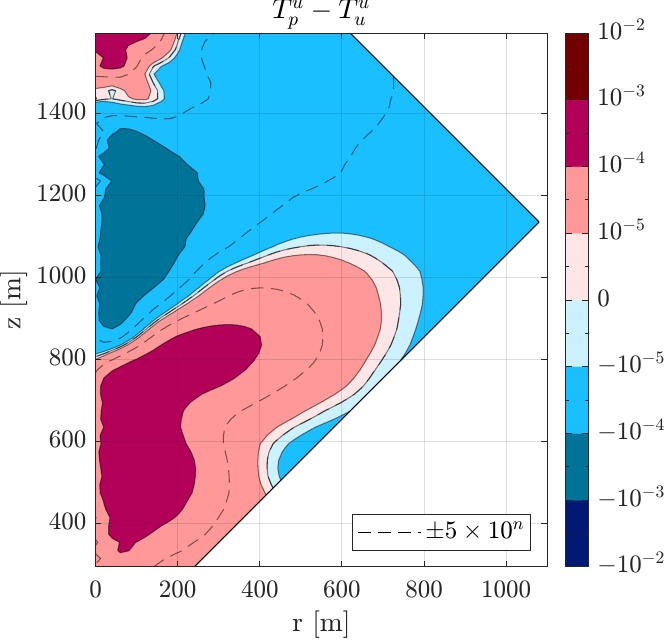}
        \caption{Interspace transfer of $\du'^2$, including the advective part
        and pressure contribution}
        \label{TuTpDiffFig}
     \end{subfigure}
     \caption{Spherical-averaged terms in the kinetic energy budget.}
    \end{figure}

Vertical stratification of the atmosphere inherently leads to the
inhomogeneity of turbulence. This includes altitude-dependent
dissipation and buoyancy forcing (see Fig.~\ref{uColormapsWE}), with
the latter changing sign just below the cloud layer at scales below 100m.
The difference between these two quantities represents an energy
surplus or deficit (see Fig.~\ref{WorkDissipDiffFig}) where the energy injection rate
cannot be balanced by local dissipation alone and must be
redistributed elsewhere. This imbalance drives the difference between pressure transport
$T_p^u$ and interspace transport $-T_u^u$. 
Note that under
the assumption of horizontal homogeneity, this transport occurs
almost exclusively in the vertical direction. It is negative in
regions with strong buoyancy work, as it exports energy away
from these zones, and positive in regions with low $W^u$
(cf. one-point statistics in Fig.~\ref{TuWMeanFig}). 
Inspecting Fig.~\ref{uColormaps1}, we note that $-T_u^u$ and $T_p^u$
are significant across all scales and act in opposition to one
another. Their similar magnitude and opposite sign makes it convenient to
combine them into one term in further analysis.
Such counter-balance
is also reminiscent of a similar behavior for one point statistics in 
atmospheric convection~\citep{McBean1975}, as
confirmed by
Fig.~\ref{TuTpMeanFig} and could be attributed to fluctuating convective updrafts.
These two fields simultaneously change sign near the cloud base 
(around $800$~m, cf. Fig.~\ref{qlMeanFig}), sharply distinguishing this region. 
However, $-T_u^u$ and $T_p^u$ do not fully balance each other out, as shown in 
Fig.~\ref{TuTpDiffFig}. We observe that the interspace transport $-T_u^u$ 
dominates at small and moderate scales, transporting (see Fig.~\ref{uColormaps1}) 
TKE into the cloud layer from below and out of the cloud layer from above. 

\FloatBarrier

\FloatBarrier
\subsection{Estimation of the total scale-by-scale turbulence energy budget}
\label{secLogPlots}
This section aims to identify distinct turbulent energy balance regimes 
that could be used in future subgrid-scale modeling. 
With this goal, we focus mainly on the TKE budget, which is well resolved and explicitly
available via the second-order velocity structure function.

The resolved TPE budget is far more difficult to express with a simple, reliable
balance due to approximation and modelling uncertainties.
We estimate that budget with the buoyancy second-order structure function 
$\delta b'^2$, normalized by $N^2$ at the center of each scale ($\ol{N^2}$). The latter
is usually assumed constant, but to take its inhomogeneity (dependence on height)
into account we compute it
directly from the profile in Fig.~\ref{refProfiles}, for each scale separately. 
This choice matches the work energy exchange at small scales, 
$W^u(r\to0) = W^b(r\to0)/\ol{N}^2$, but serves as a relatively crude estimate at 
larger scales. 
As shown in section~\ref{secColormaps}, the buoyancy budget suffers additionally from large 
SGS terms contribution, which makes it more difficult to identify specific
processes. For that reason, we focus on more instructive, directly resolved 
terms, notwithstanding of the drawback that they are often insufficient to close the budget on 
their own. Despite the limitations, the TPE budget provides a valuable
context for formulating TKE balances and their phenomenological interpretations.

In the formulation of both budgets (Eq.~\ref{KHMHeq}), all the 
terms besides the time derivative are placed on the right-hand side. 
Consequently, positive values indicate an energy gain at scales smaller than 
$r$ driven by the corresponding physical process, whereas negative values denote 
an energy loss. 
To examine turbulence energy balance regimes, we inspect in  
Fig.~\ref{logTermsFig} constant-$Z$ slices of fields selected from colormaps 
(Fig.~\ref{uColormapsWE}, \ref{uColormaps1}, \ref{bColormaps1}, \ref{bColormaps2}).

\subsubsection*{Buoyancy work region (Z = 295 -- 455m):}
Starting at $Z = 295~\text{m}$ 
we observe high buoyancy work at scales large enough to be affected by surface
heating (note its influence also in Fig~\ref{TuWMeanFig} $z\approx100~\text{m}$). 
As smaller and smaller scales are considered, that contribution decreases
being gradually replaced by inter-scale transfer, which grows towards small scales.
That transition is found to respect the following balance equation:
\begin{equation}
    4\epsilon^u \approx W^u -S_3^u
    \label{balanceWS}
\end{equation}
valid in the entire range of resolved scales and altitudes until approximately 
$Z = 455~\text{m}$. 
The joint contribution of pressure and interspace transfer is about
an order of magnitude smaller than the dominant balance (\ref{balanceWS}), however
it keeps increasing with altitude.

In the case of TPE a corresponding balance is:
\begin{equation}
    4\epsilon^b \approx Q^b -S_3^b
\end{equation}
which remains valid for scales smaller than about $r = 100\text{m}$ and altitudes below 
$Z = 495\text{m}$, until $T_u^b$, $W^b$ grow large together with SGS contributions.

\subsubsection*{Convection region (Z = 495 --- 775m):}
As we move upwards further away from the surface, we note the buoyancy work dropping
significantly at the small scales, and the joint contribution of pressure and 
interspace transfer taking its place. This leads to a smooth transition between 
balance~(\ref{balanceWS}) and:
\begin{equation}
    4\epsilon^u \approx -S_3^u + T_p^u - T_u^u.
    \label{balanceTS}
\end{equation}
The boundary between the two regimes is somewhat arbitrarily set at $Z = 475\text{m}$ where
$W^u$ and $T_p^u-T_u^u$ are approximately equal at small scales.
We also note that the new regime does not cover all the scales but rather
ones below the vertical velocity correlation length scale 
$L_{xy}^w \approx 110\text{m}$ (see Fig.~\ref{LFig}).

At the largest scales the old balance~(\ref{balanceWS}) seems to be maintained.
The strong buoyancy work in this scale range remains due to
phase changes occurring above the lifting condensation level ($Z \approx 800\text{m}$). 
This manifests clearly by a jump in ${Q^b}$ and resulting sign change of 
$-T_u^b$ and $-S_3^b$. 
We also note a somewhat inconsistent behavior of $W^u$ and $W^b$.
Such troublesome coupling of TKE and TPE equations at this altitude 
is a consequence of varying $\overline{N^2}$
(see Fig.~\ref{refProfiles}).

The described change of regime from (\ref{balanceWS}) to (\ref{balanceTS}) is likely an indicator of convective updrafts 
slowing down (more TKE is brought from below than released upwards) as most of
their TPE has already been spent. At $Z = 755\text{m}$ we note negative buoyancy
work, which introduces a direct capping to previously accelerated updrafts.

\subsubsection*{Lower in-cloud region (Z = 795 --- 1335m):}
Inside the cloud layer, which begins around $800~\text{m}$
(Fig.~\ref{qlMeanFig}), local phase changes act across all scales,
introducing a large energy surplus through the buoyancy source term ($Q^b$).
Notably, it is nearly uniform in $r$, covering also small scales
in contrast to conventional turbulence studies that typically focus on flows with 
large-scale forcing.
Although most of this extra
energy is immediately destroyed in TPE form ($|4\varepsilon^b| >
|W^b|$), the remaining surplus is so high that it is actively opposed
by all the other significant terms in the TPE budget. The opposition is
manifest through increased dissipation, more transfer to TKE via
$W^b$ (especially at small scales), and a sign reversal in both $-T^b_u$ and $-S_3^b$ 
across all scales. Such small-scale $-S_3^b$ inversion could be troublesome for
the Smagorinsky subgrid-scale model used in the simulation, which is
incapable of capturing backward energy transfer (backscatter). This
limitation is likely the reason why the SGS terms in the buoyancy
budget (Fig.~\ref{bColormaps1}) appear significantly larger than those
in the velocity equation. 

The sudden rise of small-scale $W^u$ immediately invalidates the turbulence
balance established below the cloud region. Besides the smallest scales where
$S_3^u$ has its peak, $W^u$ exceeds both $S_3^u$ and $T_p^u - T_u^u$
by about an order of magnitude. This can be expressed with the rough balance
\begin{equation}
    4\epsilon^u \approx W^u
    \label{balanceW}
\end{equation}
and its more accurate extension:
\begin{equation}
    4\epsilon^u \approx W^u -S_3^u + T_p^u - T_u^u.
    \label{balanceWST}
\end{equation}
which handles inverse interscale transfer very well.
The former equation (\ref{balanceW}) is sufficient where 
the inverse transfer is opposed by interspace transport and pressure, as is the
case in the range $835\text{m} < Z < 915\text{m}$. 
The lower in-cloud regime reaches up to $Z = 1335\text{m}$, where the
forward interscale transfer is recovered. 

Further up, 
a significant unsteady contribution (see Fig.~\ref{uColormaps1}) and 
piling up of all the other terms (including SGS model influence)
makes it difficult to propose a simple turbulence balance (see $Z = 1595\text{m}$).

A possible physical picture for the characteristics described above in this section is a set
of initially small, numerous, and newly formed updrafts. These
emerging structures have high buoyancy rather than velocity which
they have not yet had time to build up. As they rise, they carry
buoyancy upward. Naturally, they also expand as they rise, which leads
to merging processes and an inverse TPE cascade. As these updrafts
grow larger and mature at slightly higher altitudes, they develop a
forward TPE cascade. This development can be traced by looking at the
small-scale, red $-S_3^b$ region in Fig.~\ref{bColormaps1} and is also
notable at $Z = 1075~\text{m}$.

The presence of these new updrafts also shows up in the TKE budget
at $Z = 855~\text{m}$ through negative values of $-{T_u^u}$, 
which covers scales below about $100~\text{m}$. 
This scale range matches well with the integral length scale based on 
the vertical velocity, $L^w_{xy}$ (see Fig.~\ref{LFig}). 
Above this scale, but still below
the horizontal velocity length scale $L_{xy}^u$, we see an inverse
inter-scale transfer ($-{S_3^u} < 0$).
Structures made of horizontal correlations lose energy to their larger equivalents
at the same altitude. That transfer is fueled from below, by inter-space transport
possibly in a form of uncorrelated assembly of updrafts.
This way we can see turbulence energy moving up in $Z$ and up in scales at 
the same time.
About $Z = 1075~\text{m}$ the dynamics at the same scales change as they 
are increasingly affected by purely in-cloud phenomena.
The buoyant forcing significantly exceeds local dissipation and that surplus
is directly passed to both larger scales in the same place and same-sized 
structures upwards.
At the same time, the updrafts sustain a forward TKE cascade within
their cores ($r < L^w_{xy}$), despite inverse TPE cascade at the same scales. 
This is probably because large
velocities characterize the more mature updrafts (which have already
developed this forward cascade) acting as a filter on the overall
statistics. There could also be a contribution from surface-born
updrafts that possessed enough kinetic energy to penetrate the adverse
buoyancy zone below, importing mature turbulent structures from lower
altitudes. This nonlocal transport is not mirrored in the buoyancy
budget, which strictly depends on the local background stratification
and immediate phase changes. At a higher altitude ($Z =
1075~\text{m}$), the scale range of negative $-{T_u^u}$ is larger,
which is consistent with the upscale mergers described above. On the
other hand, the inverse inter-scale transfer covers a similar range of
scales as it does at $Z = 855~\text{m}$.

We also notice that the buoyancy characteristics at $Z = 855~\text{m}$
and $Z = 1075~\text{m}$ are less dependent on the scale $r$. 
This is most likely due to the weaker buoyancy below the cloud layer,
which again acts like a filter on the data. Consequently, the
contribution of different altitudes introduced via large scales is
weak.

\subsubsection*{Upper in-cloud region (Z = 1335 -- 1595m):}
In the upper in-cloud region, about $z = 1595~\text{m}$,
the mature updrafts begin to decelerate,
despite ongoing phase changes. More 
energy is advected from the regions below than is transported upward. 
Consequently, because all the mergers that were going to happen have already 
occurred, the forward cascade is recovered for both the TKE and TPE budgets.
A significant contribution of the unsteadiness term is also worth noting in the small
and intermediate scales. 
The weaker inter-scale transfer and negative inter-space transfer observed 
at the largest scales are due to the contribution of the lower in-cloud region.

\begin{figure}[h!]
     \centering
    \begin{subfigure}[b]{\cmSize\textwidth}
         \centering
         \includegraphics[width=\textwidth]{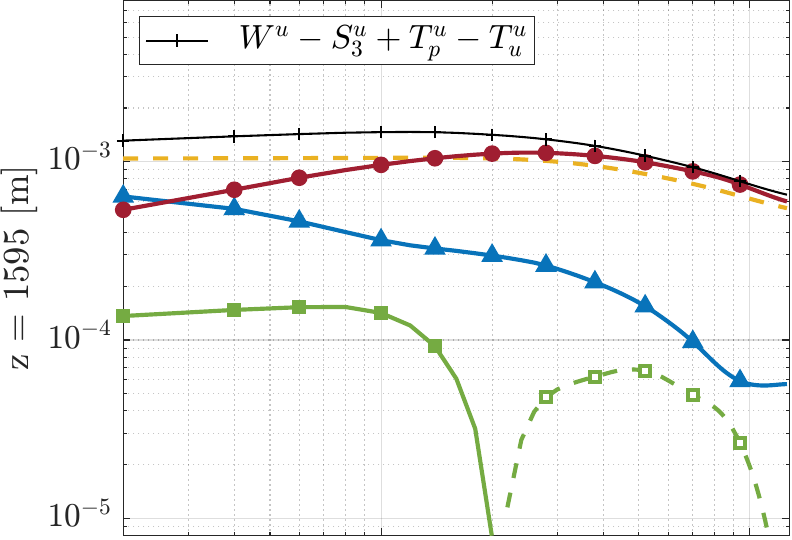}
     \end{subfigure}
     \hfill
     \begin{subfigure}[b]{\cmSize\textwidth}
         \centering
         \includegraphics[width=\textwidth]{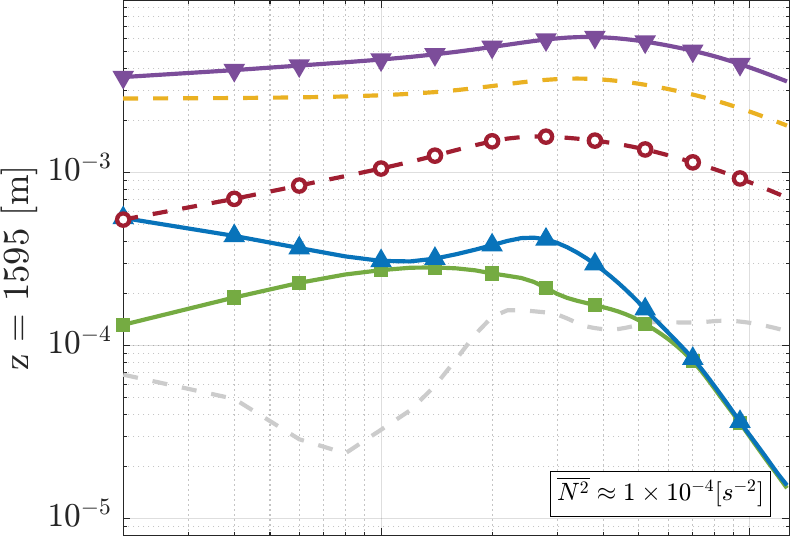}
     \end{subfigure}
    \begin{subfigure}[b]{\cmSize\textwidth}
         \centering
         \includegraphics[width=\textwidth]{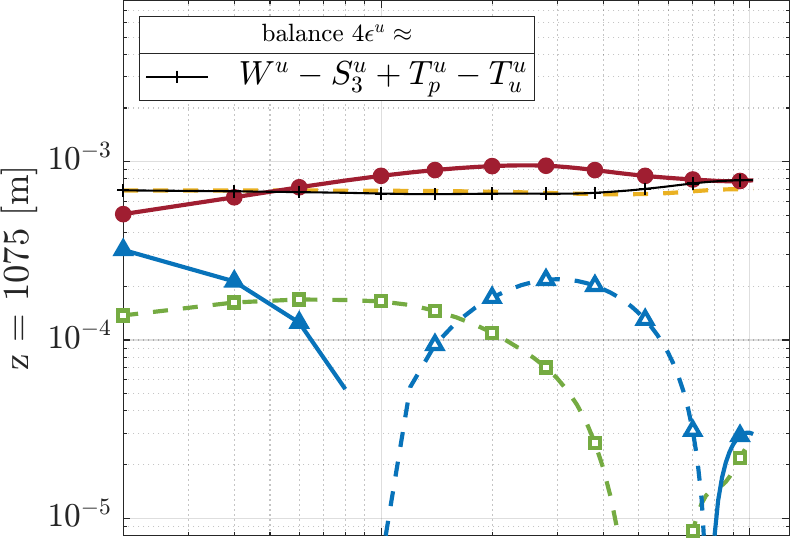}
     \end{subfigure}
     \hfill
     \begin{subfigure}[b]{\cmSize\textwidth}
         \centering
         \includegraphics[width=\textwidth]{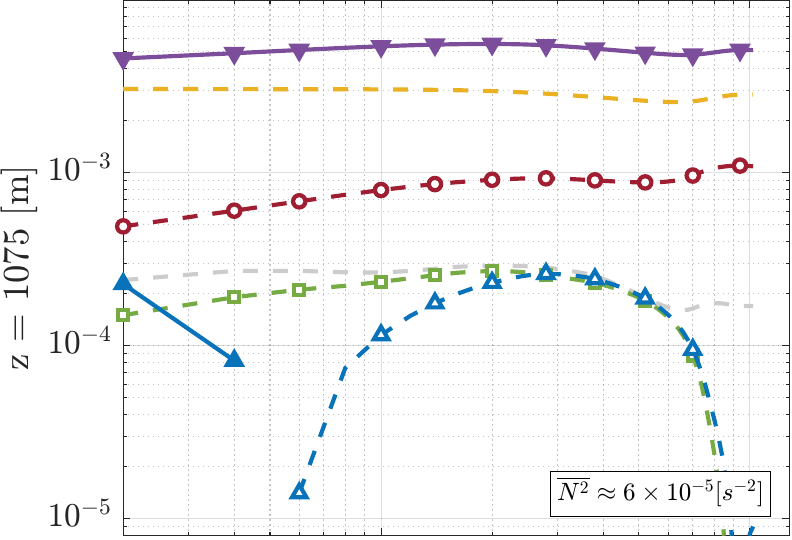}
     \end{subfigure}
    \begin{subfigure}[b]{\cmSize\textwidth}
         \centering
         \includegraphics[width=\textwidth]{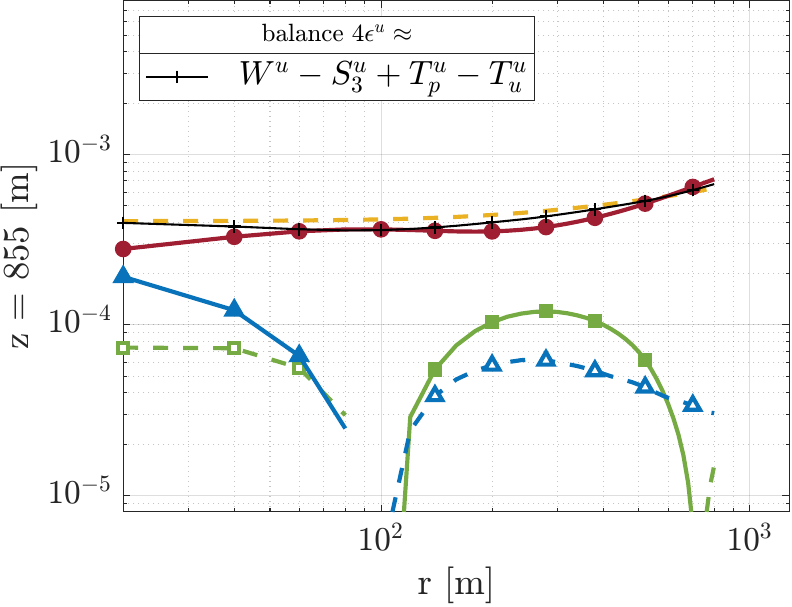}
     \end{subfigure}
     \hfill
     \begin{subfigure}[b]{\cmSize\textwidth}
         \centering
         \includegraphics[width=\textwidth]{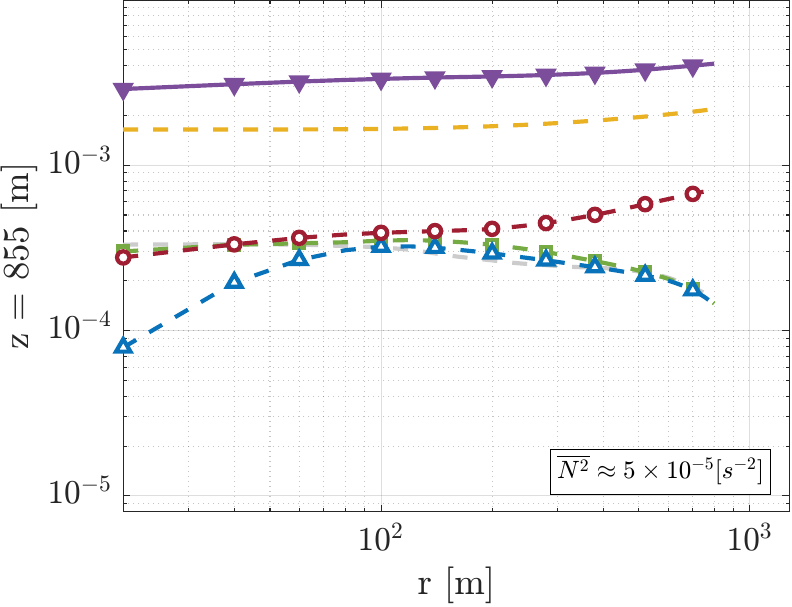}
     \end{subfigure}
\end{figure}
\begin{figure}
    \centering
    \ContinuedFloat
    \begin{subfigure}[b]{\cmSize\textwidth}
         \centering
         \includegraphics[width=\textwidth]{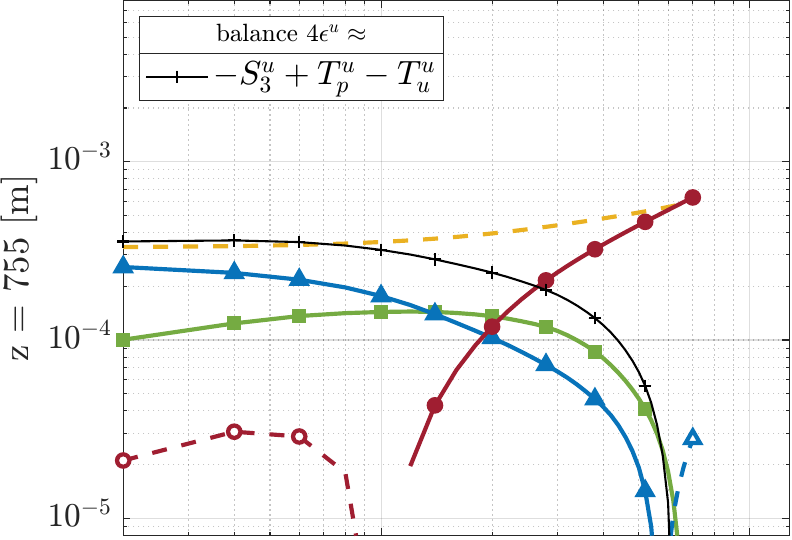}
     \end{subfigure}
     \hfill
     \begin{subfigure}[b]{\cmSize\textwidth}
         \centering
         \includegraphics[width=\textwidth]{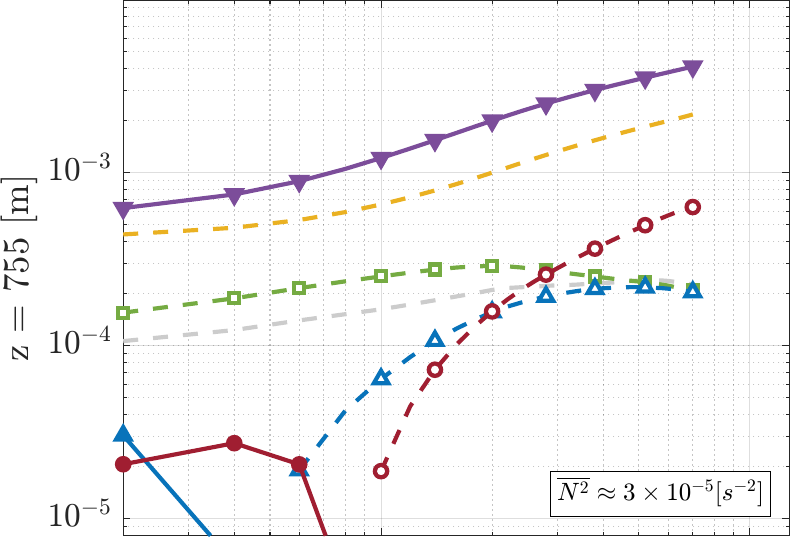}
     \end{subfigure}
    \begin{subfigure}[b]{\cmSize\textwidth}
         \centering
         \includegraphics[width=\textwidth]{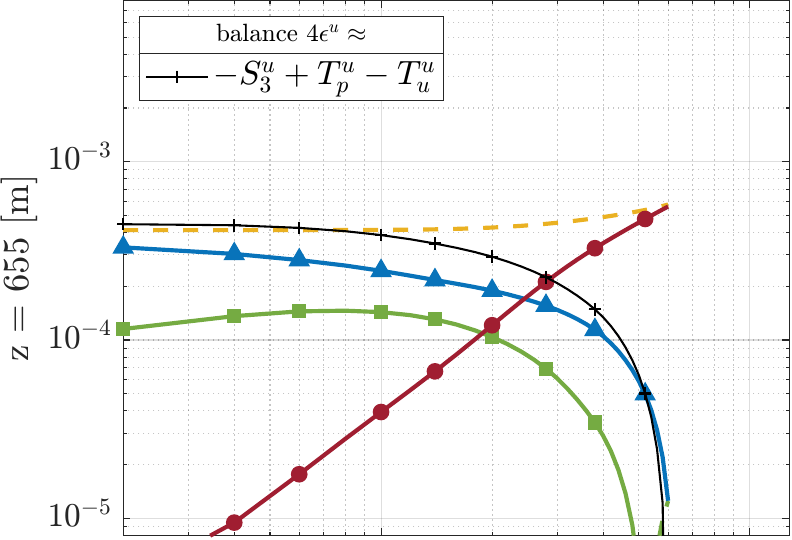}
     \end{subfigure}
     \hfill
     \begin{subfigure}[b]{\cmSize\textwidth}
         \centering
         \includegraphics[width=\textwidth]{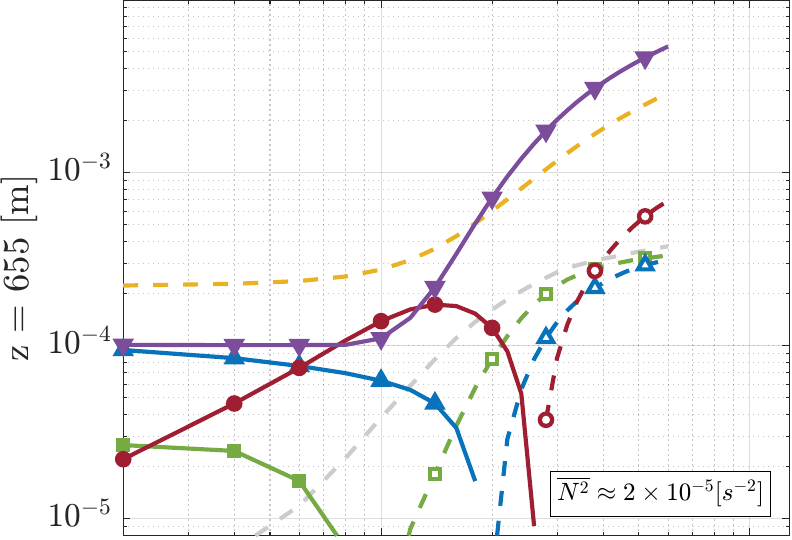}
     \end{subfigure}
    \begin{subfigure}[b]{\cmSize\textwidth}
         \centering
         \includegraphics[width=\textwidth]{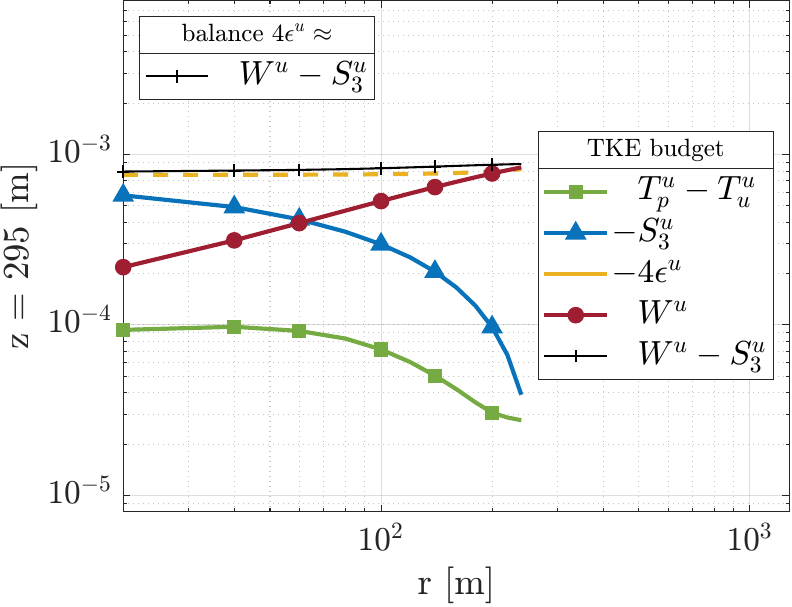}
     \end{subfigure}
     \hfill
     \begin{subfigure}[b]{\cmSize\textwidth}
         \centering
         \includegraphics[width=\textwidth]{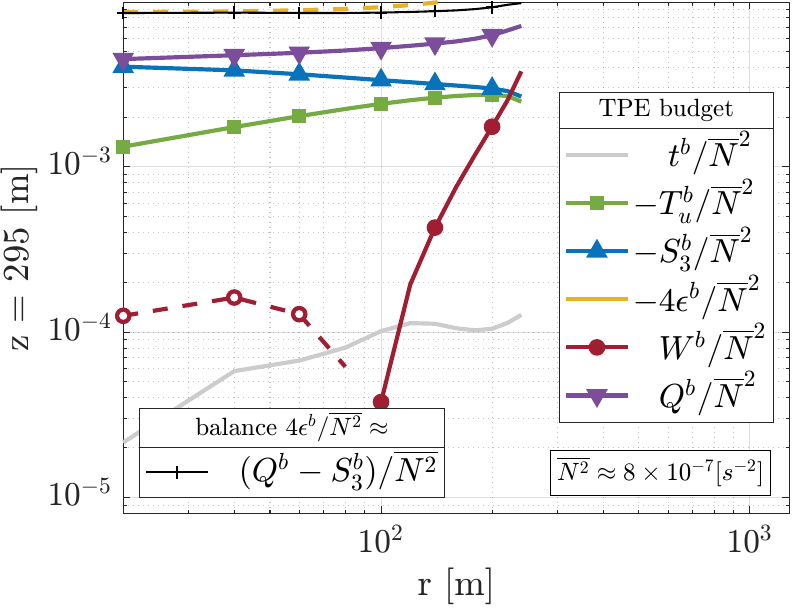}
     \end{subfigure}
     \caption{Selected terms of turbulent kinetic (left column) and estimation of turbulent 
     potential energy 
     (right column) in logarithmic scale. Dashed lines denote intervals with negative values.
     Each column has its own legend, located in the bottom subfigure.}
     \label{logTermsFig}
\end{figure}

\FloatBarrier

\section{Summary and Conclusions}
\label{secConclusions}

This study investigate the scale-by-scale budgets of velocity and buoyancy 
second-order structure functions using 
high-resolution LES data of a classic trade-wind 
cumulus layer (BOMEX). 
The velocity 
structure functions provide an explicit representation of the turbulence kinetic energy 
budget, while the buoyancy structure functions have been used to estimate the scale-by-scale 
budget of the available turbulence potential 
energy. The simulation utilize 
a $10~\text{m}$ grid resolution with a 
Smagorinsky subgrid-scale model, employing the anelastic approximation 
to account for vertical stratification and moist thermodynamics to fully 
capture phase changes. 
By examining each term of the TKE and TPE equations scale-by-scale
this work has identified turbulence balances
beyond the assumptions of homogenous isotropic 
turbulence theory.

The features of atmospheric turbulence were split into four distinct regions,
in three of which we have successfully identified approximate TKE balances remaining the
same across a wide range of scales.
In the case of TPE, the uncertainties related to subgrid-scale modeling not allows
to capture the behavior in simple, reliable balance formula, however, the TPE budget does 
provide a valuable context for interpretation of TKE dynamics.

A concise summary of the findings is gathered in Table~\ref{tab:regimes} and briefly
commented below. For a more detailed description, the reader is referred to Section~\ref{secLogPlots}.

\begin{table}[ht!]
\centering
\renewcommand{\arraystretch}{1.25}
\begin{tabular}{||l|l|l||}
\hline
regime name & altitude range & dominant TKE balance \\ \hline\hline
buoyancy work   & $295\phantom{0}-\phantom{0}455~\text{m}$    & $4\epsilon^u \approx W^u - S_3^u$  \\
convection      & $495\phantom{0}-\phantom{0}775~\text{m}$    & $4\epsilon^u \approx \phantom{W^u}- S_3^u + T_p^u - T_u^u$ \\
lower in-cloud  & $795\phantom{0}-1335~\text{m}$   & $4\epsilon^u \approx W^u - S_3^u + T_p^u - T_u^u$  \\
upper in-cloud  & $1335-1595~\text{m}$  & complex / inconclusive \\ \hline
\end{tabular}
\caption{Summary of the identified regimes. For definitions of notation see Section~\ref{secTerms}.}
\label{tab:regimes}
\end{table}

At the lowest altitudes studied, surface heating manifests through increased buoyancy work, 
mostly at large scales. As the scale becomes smaller, this effect decreases and gives way to 
inter-scale transfer. At higher altitudes, it is smoothly replaced by spatial and pressure 
transport. 
Before in-cloud phase changes enter the budget, these processes form a balance with buoyancy 
work at small and intermediate scales.

Above the Lifting Condensation Level, phase changes introduce a buoyancy source 
and resulting buoyancy work across all resolved scales. This leads to significant 
inverse inter-scale transfer, spanning all scales for TPE and intermediate scales 
for TKE. For TKE, this intermediate range lies between the vertical velocity integral 
length scale ($\approx 100\,\mathrm{m}$) and the horizontal velocity scale 
($\approx 700\,\mathrm{m}$). This is accompanied by spatial and pressure transports 
of similar magnitude, such that all three terms counteract the local energy surplus 
($W^u - 4\epsilon^u > 0$), which peaks around $Z \approx 1075\,\mathrm{m}$ at 
intermediate scales.

In the upper cloud layer, despite ongoing phase changes, positive contributions 
from interscale transfer as well as spatial and pressure transports are recovered, 
which is more typical of sub-cloud regions. However, due to significant 
unsteadiness and the growing impact of secondary terms, no simple and reliable 
balance could be established across scales.

The numerical results show satisfactory agreement with a previous empirical study based on 
airborne measurements \citep{Nowak2025}, which focused on sub-cloud altitudes. 
This agreement is confirmed when applying the observational methodology to the LES data 
by comparing aircraft tracks against one-dimensional subsamples. Combined with the low 
statistical standard errors of the computed terms, this agreement validates the physical 
consistency and reliability of the results.%

Ultimately, the findings of this study demonstrate that atmospheric 
turbulence equilibria go beyond the forward energy cascade paradigm. 
Rather, they emerge from an interplay between spatial and pressure transport,
small-scale forcing, and direction-altering energy cascades.

\section*{Data Availability Statement}
The data that support the findings of this study are available from the corresponding author upon reasonable request.

\section*{Funding Statement}
This work was funded by the European Union (ERC, NoStaHo, 101054117). 
Views and opinions expressed are, however, those of the authors only and do not necessarily 
reflect those of the European Union or the European Research Council. Neither the European 
Union nor the granting authority can be held responsible for them. \\

\section*{Conflict of Interest Statement}
The authors declare no conflicts of interest.

\section*{Acknowledgements}
Marta Wacławczyk was Visiting Professor at the University of Lille in the 
summer of 2026 and Szymon Malinowski was Visiting Professor at Ecole Centrale de Lille 
in the summer of 2023 and at the University of Lille in the summer of 2024. 
They both participated in the Lille Turbulence Programmes of 2023, 2024, 2025 and 2026. 
Paweł Jędrejko participated in the Lille Turbulence Programmes of 2025 and 2026.

\noindent This research was carried out with the support of the Interdisciplinary 
Centre for Mathematical and Computational Modelling University of Warsaw 
(ICM UW) under computational allocations no g100-2225 and g102-2487.

\bibliography{references}

\appendix
\section{Anelastic approximation}
\numberwithin{equation}{section}
\label{Appendix}

 Basic assumptions  behind the anelastic approximation can be explained 
 by introducing the concept of a parcel of air, which is displaced from its equilibrium position at $z$ to $z+D$. 
 The parcel experiences a change in its temperature, density and pressure while it raises and the difference between the actual parcel properties and the background state are $ \rho' = \rho-\rho_0$, $ p' = p-p_0$, $T'= T- T_0$.     
 The anelastic approximation assumes that 
 these differences (perturbations) are small compared to the static values
 \begin{align}
 	\frac{\rho'}{\rho_0(z)}  \ll 1, \quad 	
 	\frac{p'}{p_0(z)} \ll 1, \quad 	\frac{T'}{T_0(z)} \ll 1
 	\label{eq:assumption1A}
 \end{align}
 and that the parcel displacement $D$ may be comparable to the density scale height $H_\rho$ defined
 in Eq.\ (\ref{def:H})
 \begin{equation}
 	\frac{D}{H_\rho} \sim {\cal O}(1).
 	\label{eq:assumption2}
 \end{equation}

\subsection{Continuity equation}
 The decomposition into the background state and the perturbation $ \rho'$
 is now introduced into the continuity equation for a compressible fluid,
 which leads to
 \begin{equation}
 	\frac{1}{\rho_0}\frac{D  \rho'}{D t} 
 	+
 	w \frac{1}{\rho_0}\frac{{\rm d} \rho_0(z)}{{\rm d} z}  + 
 	\left( 1+\frac{ \rho'}{\rho_0} \right) \frac{\partial u_i}{\partial x_i} = 0.
 	\label{eq:conti1}
 \end{equation}
 where $D/Dt=\partial/\partial t + u_i \partial/\partial x_i $.
 
 The first term in the equation above describes the change of the perturbation density along the  parcel track. 
 This term is of order
 \begin{equation}
 	\frac{1}{\rho_0}\frac{D  \rho'}{D t} \sim \frac{ \rho'}{\rho_0 \tau} \sim \frac{\rho'}{\rho_0}  \frac{{\cal W}}{D}
 	\label{eq:term1}
 \end{equation}
 where $\tau = D/{\cal W}$ is the timescale and  ${\cal W}$ is the characteristic scale of the vertical velocity.
 The order of the second term in Eq.\ (\ref{eq:conti1}) can be determined using  the definition (\ref{def:H})
 \begin{equation}
 	w \frac{1}{\rho_0}\frac{{\rm d} \rho_0(z)}{{\rm d} z} \sim 
 	\frac{{\cal W}}{H_\rho}.
 	\label{eq:term2}
 \end{equation}
 Due to assumptions (\ref{eq:assumption1A}) and (\ref{eq:assumption2}),
 this term should be much larger than the first one
 and hence must be balanced by the third LHS term in Eq.\ (\ref{eq:conti1}).
 It then follows 
 that the leading order terms in Eq.\ (\ref{eq:conti1}) are
 \begin{equation}
 	w \frac{1}{\rho_0}\frac{{\rm d} \rho_0(z)}{{\rm d} z} + 
 	\frac{\partial u_i}{\partial x_i} = 0, 
 	\label{eq:conti2}
 \end{equation}
 which can be rewritten, using  (\ref{def:H}),
 in the following equivalent forms 
 
 \begin{equation}
 	\frac{\partial \rho_0 u_i}{\partial x_i} = 0
 	\quad \text{or}
 	\quad \frac{\partial u_i}{\partial x_i} = \frac{w}{H_\rho}
 	\label{eq:conti3}
 \end{equation}
 
 The assumption (\ref{eq:assumption2}) represents the key difference between the anelastic and Boussinesq approximations. The latter assumes that
 $ {\partial u_i}/{\partial x_i} =0 $, or equivalently, that the density of a fluid parcel does not change as it rises. 
 The anelastic and Boussinesq approximations become equivalent when
 the density scale height 
 $H_\rho$ becomes large, when compared to the parcel displacement $D$.
 
\subsection{Momentum and energy transport for dry air}
 
 The same analysis will be applied to the 
 momentum and energy equations
 \begin{align}
 	\rho \frac{D u_i}{D t} =& -\frac{\partial p'}{\partial x_i} 
 	-\rho' g \delta_{i3} + \rho f \epsilon_{ij3} 	\left({u}_j-{G}_{j}\right)+
 	\mu  \frac{\partial^2 u_i}{\partial x_j \partial x_j} ,
 	\label{eq:mom} \\
 	\rho c_p \frac{D { \theta}}{D t} =& 
 	\lambda \frac{\partial^2 \theta}{\partial x_j \partial x_j} +
 	\dot{Q} \left(\frac{p_{00}}{p}\right)^{R/c_p}
 	\label{eq:en}
 \end{align}
 where  $f= 2\Omega \sin(\phi)$ is the Coriolis parameter, $\phi$ is the latitude and $\textbf{G}$ is the geostrophic velocity, $\delta_{ij}$
 is the Kronecker delta,  $\epsilon_{i3k}$ is the Levi-Civita symbol
 and $\mu$ is the dynamic viscosity.
 We subtracted  the hydrostatic  and geostrophic equilibrium equation
 $$
 0=-\frac{\partial p_0}{\partial x_i}  -\rho_0 g \delta_{i3} + \rho f \epsilon_{ij3} G_j  $$
 to
 replace the pressure $p$  by the perturbation pressure in the momentum equation (\ref{eq:mom}).
 Further,
 $\theta$ is the potential temperature defined as
 \begin{equation}
 	\theta = T \left(\frac{p_{00}}{p}\right)^{R_d/c_p},
 	\label{def:potTA}
 \end{equation}
 where $p_{00}$ is the reference pressure, $R_d=c_p - c_v$ is the specific gas constant for dry air and
 $c_p$ and $c_v$ are, respectively, the specific heat at constant pressure and at constant volume,
 $\lambda$ is the thermal conductivity coefficient and
 $\dot{Q}$ is the diabatic heating rate per unit volume, which includes the latent heating, cf.\ \citep{Kurowski2014}.
 The above equations are next divided by the reference density $\rho_0$.
 We also make use of the assumption (\ref{eq:assumption1A}), which implies that   $ \rho'/\rho_0$ is
 small compared to unity, so we can neglect it  everywhere, apart from the gravity force term. As a result we obtain the 
 following approximations of the momentum and energy equations
 \begin{align}
 	\frac{D u_i}{D t} =& -\frac{\partial}{\partial x_i} \left(\frac{ p'}{\rho_0}\right) +g\left(\frac{p'}{\rho_0 g H_\rho}-\frac{ \rho'}{\rho_0}\right) \delta_{i3}+   \epsilon_{ij3} f (u_j-G_j)
 	+ 	\nu  \frac{\partial^2 u_i}{\partial x_j \partial x_j}, \label{eq:momentum}\\
 	\frac{D { \theta}}{D t} =& 
 	\kappa  \frac{\partial^2 \theta}{\partial x_j \partial x_j}+
 	\frac{1}{c_p \rho_0}\dot{Q} \left(\frac{p_{00}}{p_0}\right)^{R/c_p},
 	\label{eq:energy}
 \end{align}
 where $\nu=\mu/\rho_0$, $\kappa=\lambda/c_p \rho_0$ and we used the definition of $H_\rho$ given in (\ref{def:H}).
 Based on this form, \cite{Bannon}, proposed suitable  approximation of $ \theta'$. The
 main steps of the derivation are outlined below.
 As follows from the ideal gas law, the assumption (\ref{eq:assumption1A}) 
 and the definition of the potential temperature (\ref{def:potTA}) the perturbations
 $\theta'$ are expressed as
 \begin{equation}
 	\frac{\theta'}{\theta_0} \approx \frac{p'}{\gamma p_0}-\frac{\rho'}{\rho_0},
 	\label{eq:theta0}
 \end{equation}
 where $\gamma=c_p/c_v$. In the ideal gas $\gamma p_0 = c^2 \rho_0$, where $c$
 is the speed of sound. If the background state is adiabatic, then
 from the definition of $c$ we obtain
 \begin{equation}
 	c^2 = \left( \frac{\partial p}{\partial \rho} \right)_s = \left( \frac{\partial p}{\partial z} \right)_s \left( \frac{\partial \rho}{\partial z} \right)_s^{-1}=
 	g H_\rho
 \end{equation}
 where the last equality follows from the hydrostatic balance equation and the definition of $H_\rho$.
 With this $\gamma p_0$ in Eq.\ (\ref{eq:theta0}) can be replaced by $\rho_0 g H_\rho$
 \begin{equation}
 	\frac{\theta'}{\theta_0} \approx \frac{p'}{\rho_0 g H_\rho}-\frac{\rho'}{\rho_0}.
 	\label{eq:theta1}
 \end{equation}
 This form is identical to the bracketed term in Eq.\ (\ref{eq:momentum}).
 
 \subsection{Formulation of moisture and phase changes}
 
 To account for moisture effects, \cite{Bannon} expressed the density $\rho$ in terms of
 the density of dry air $\rho_d$ and the mixing ratios of water vapour and liquid water,
 $q_v$ and $q_l$, respectively
 \begin{equation}
 	\rho = \rho_d\left(1+q_v + q_l\right)
 \end{equation}
 Its linearised form, under the additional assumptions that $q_v \ll 1$ and $q_l \ll 1$ is
 \begin{equation}
 	\frac{\rho'}{\rho_0} = \frac{\rho_d'}{\rho_0} +q_v + q_l. \label{eq:moist_dens}
 \end{equation}
 Above, the dry hydrostatic reference state was chosen, i.e. $\rho_{0}=\rho_{d0}$
 and $p_0=\rho_{d 0} R_d T_0$.
 The pressure of the air parcel, expressed as a sum of partial pressures of the dry air and water vapour
 reads
 \begin{equation}
 	p = \rho_d R_d T \left(1+ \frac{R_v}{R_d} q_v\right).
 	\label{eq:moiststate}
 \end{equation}
 and its linearised form is
 \begin{equation}
 	\frac{p'}{p_0} = \frac{\rho'_d}{\rho_0} + \frac{T'}{T_0} + \frac{R_v}{R_d} q_v.
 	\label{eq:moiststate_lin}
 \end{equation}	
 The momentum equation (\ref{eq:mom}) takes the following form
 \begin{equation}
 	\frac{D u_i }{D  t} = -\frac{\partial}{\partial x_i} \left(\frac{ p'}{\rho_0}\right) +g\left(\frac{p'}{\rho_0 g H_\rho}-\frac{ \rho_d'}{\rho_0} -q_v -q_l  \right) \delta_{i3} +  f\epsilon_{ij3} (u_j-G_j)+\nu \frac{\partial^2 u_i}{\partial x_j \partial x_j}, \label{eq:momentum1}
 \end{equation}
 where the ratio $\rho'/\rho_0$ was expressed in terms of $\rho_d'/\rho_0$, with the use of
 Eq.\ (\ref{eq:moist_dens}).	
\cite{Bannon} proposed to express the perturbation potential temperature as
\begin{equation}
	\frac{\theta_v'}{\theta_0} \approx \frac{p'}{\rho_0 g H_\rho}-\frac{ \rho_d'}{\rho_0} -q_v -q_l
	\approx \frac{c_v}{c_p}\frac{p'}{p_0}-\frac{ \rho_d'}{\rho_0} -q_v -q_l,
\end{equation}
where it was assumed that $\rho_0 H_\rho g = {c_p}/{c_v} p_0$ as it would be in adiabatic reference state.

Substituting for $-{ \rho_d'}/{\rho_0}$ the linearised equation of state (\ref{eq:moist_dens}) for the moist air
we arrive at
\begin{equation}
	\frac{\theta_v'}{\theta_0} \approx \frac{T'}{T_0}
	- \frac{R}{c_p}\frac{p'}{p_0}-\left(1 -\frac{R_v}{R_d} \right)q_v -q_l,
\end{equation}
which is the linearised definition of the virtual potential temperature $\theta_v$ 
\begin{equation}
	\theta_v = T \left(\frac{p_{00}}{p}\right)^{R/c_p} \left[1-\left(1-\frac{R_v}{R_d}\right)q_v - q_l \right] =
	\theta\left[1-\left(1-\frac{R_v}{R_d}\right)q_v - q_l \right]. \label{vptA}
\end{equation}

\FloatBarrier
\section{Numerical evaluation on a staggered grid}
\numberwithin{equation}{section}
\label{secNumericalMethod}

\FloatBarrier

\begin{figure}[ht!]
    \centering
    \begin{subfigure}[b]{0.49\textwidth}
        \centering
        \includegraphics[height=4cm]{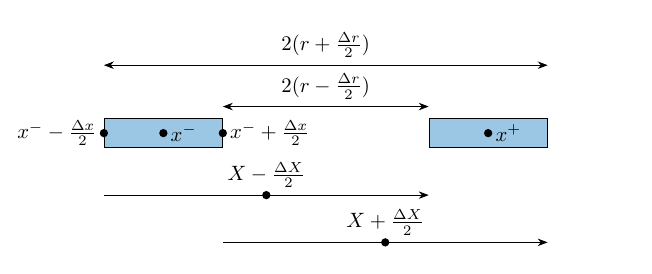}
        \caption{Two one-dimensional $x$-cells.}
    \end{subfigure}
    \hfill
    \begin{subfigure}[b]{0.49\textwidth}
        \centering
        \includegraphics[height=4cm]{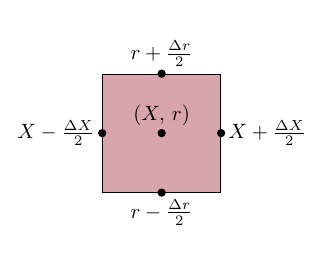}
        \caption{One two-dimensional $r-X$-cell.}
    \end{subfigure}
    \caption{Change of coordinates applied to finite volume discretization. 
    One-dimensional case for readability.}
    \label{figXiTorX}
\end{figure}

The conservative discretization of space-scale equations is achieved by evaluation of all the 
fluxes on the faces of a (6D) $r-X$ cell and then applying the same scheme as in the LES.
We relate these fluxes to values sampled on the faces of an $x$ cell referring to (\ref{xiTorX}). 
Considering steps in $X_i$ and $r_i$ directions:
\begin{equation*}
    \D X_i = \frac{\D\xip_i +\D\xim_i}{2}, \qquad 0 = \frac{\D\xip_i -\D\xim_i}{2}
\end{equation*}
what leads to $\D\xip_i = \D X_i$, $\D\xim_i = \D X_i$ and
\begin{equation*}
    0 = \frac{\D\xip_i +\D\xim_i}{2}, \qquad \D r_i = \frac{\D\xip_i -\D\xim_i}{2}
\end{equation*}
leads to $\D\xip_i = \D r_i$, $\D\xim_i = -\D X_i$.
Therefore, the 2nd order central difference of some field $\phi(r_i, X_j)$ looks as follows:
\begin{equation*}
    \frac{\phi(r_i+\tfrac{\D r}{2}, X_j) - \phi(r_i-\tfrac{\D r}{2}, X_j)}{\D r} =
    \frac{\phi(\xim_i-\tfrac{\D x}{2}, \xip_j+\tfrac{\D x}{2}) - \phi(\xim_i+\tfrac{\D x}{2}, \xip_j-\frac{\D x}{2})}{\D x}
\end{equation*}
\begin{equation*}
    \frac{\phi(r_i, X_j+\tfrac{\D X}{2}) - \phi(r_i, X_j-\tfrac{\D X}{2})}{\D X} =
    \frac{\phi(\xim_i+\tfrac{\D x}{2}, \xip_j+\tfrac{\D x}{2}) - \phi(\xim_i-\tfrac{\D x}{2}, \xip_j-\frac{\D x}{2})}{\D x}.
\end{equation*}
A graphical representation of this result is presented in Fig. (\ref{figXiTorX}).

Noting also the useful fact, that finite difference analogues of (\ref{deltaOfGrad}), (\ref{barOfGrad}) hold:
\begin{subequations}
\begin{equation}
    \d\frac{\D\phi}{\D x} = \frac{\D\d\phi}{\D X} = 2\frac{\D\ol\phi}{\D r}
    \label{gradMinusDisc}
\end{equation}    
\begin{equation}
        \ol{\frac{\D\phi}{\D x}} = \frac{\D\ol\phi}{\D X} = \frac{1}{2}\frac{\D\d\phi}{\D r}.
        \label{gradPlusDisc}
\end{equation}
\label{gradDisc}
\end{subequations}
we derive all the $r-X$ analogues of the schemes used in the LES in the similar fashion.

Unfortunately, the direct discretization  of terms from Tab. (\ref{termsTab}) 
does not lead to the closed budget,
due to additional term in a finite difference product rule.
To achieve a closed budget we take a few steps back and use closure of the governing equation, 
which holds after multiplication by a scalar.
Before the discretization, the anelastic terms are expressed as:
\begin{equation*}
    A^u = 2\d\Big(\frac{\p u_i' u_j'\rho}{\p x_j} \Big) \du_i'- S^u_3 - T^u_u,
\end{equation*}
and
\begin{equation*}
    D^u_A = 2\d\Big(\frac{1}{\rho}
    \frac{\p}{\p x_j}\big(\rho\nu S_{ij}'\big)\Big)\du_i' - (D_{XX}^u + D_{rX}^u + D_{Xr}^u + D_{rr}^u + 4\varepsilon^u) 
\end{equation*}
where we just use the values of $x$-space terms on the right-hand side.
The diffusive terms need a similar treatment and are first expressed as:
\begin{equation*}
    D_{XX} = 2\frac{\p}{\p X_j}\Big(\ol{\nu}_s S_{ij}^{X,\d}\Big)\du_i'
    + 2\ol{\nu}_s S_{ij}^{X,\d}S_{ij}^{X,\d},
\end{equation*}
\begin{equation*}
    D_{Xr} = \frac{\p}{\p X_j}\Big(\d\nu_s S_{ij}^{r,\d}\Big)\du_i'
    + \d\nu_s S_{ij}^{r,\d}S_{ij}^{X,\d},
\end{equation*}
\begin{equation*}
    D_{rX} = \frac{\p}{\p r_j}\Big(\d\nu_s S_{ij}^{X,\d}\Big)\du_i'
    + \d\nu_s S_{ij}^{X,\d}S_{ij}^{r,\d},
\end{equation*}
\begin{equation*}
    D_{rr} = 2\frac{\p}{\p r_j}\Big(\ol{\nu}_s S_{ij}^{r,\d}\Big)\du_i'
    + 2\ol{\nu}_s S_{ij}^{r,\d}S_{ij}^{r,\d}
\end{equation*}
where the second terms on the right-hand side are evaluated in the cell center and add up to minus dissipation term.

\end{document}